\documentclass[proceedings]{ccn}   
\ccndoi{10.32470/7dlktq8}  

\usepackage{amsmath,amssymb}
\usepackage{bm}
\usepackage{multirow}
\usepackage{graphicx}
\usepackage{subfigure}
\usepackage{booktabs}
\usepackage{hyperref}
\usepackage{microtype}
\usepackage{placeins} 
\usepackage[table]{xcolor}
\usepackage{float}

\title{Task-guided cross-subject latent alignment:\\a multi-encoder-decoder VAE}

\author{Angeliki Papathanasiou\textsuperscript{1}, Jascha Achterberg\textsuperscript{1}, Thomas E. Nichols\textsuperscript{2}, Rui Ponte Costa\textsuperscript{1}
\\
$^{1}$ Centre for Neural Circuits and Behaviour, Department of Physiology Anatomy and Genetics, University of Oxford \\
$^{2}$ Big Data Institute, Nuffield
Department of Medicine,  University of Oxford
}

\begin{document}

\maketitle

\begin{abstract}
\textbf{Aligning neural activity across subjects offers the promise of discovering shared computational principles and generalizable decoders. However, traditional alignment methods require shared stimuli across subjects, a constraint that limits applicability to naturalistic paradigms with limited or non-overlapping data. We introduce a Multi-Encoder-Decoder Variational Autoencoder (MED-VAE) that achieves cross-subject alignment without shared stimuli by anchoring representations to a common scaffold provided by a pretrained ANN.  Using the Natural Scenes Dataset, we show that MED-VAE creates common latent spaces with superior semantic organisation, achieving higher cross-subject alignment than common methods while maintaining robust generalisation to held-out stimuli where traditional methods degrade.
Reconstructing from these common spaces back to each subject's original neural space, MED-VAE preserves equal stimulus-driven signal in its cross-subject latent space.
Finally, we show that this superior alignment directly enables cross-subject neural prediction, as demonstrated via cross-subject image decoding.
In summary, we introduce a framework to identify generalisable common subspaces for cross-subject predictions and downstream tasks, demonstrated here for visual cortex responses to static images. 
}
\end{abstract}

\section{Introduction}
\label{sec:intro}

Neural responses to identical stimuli vary substantially across individuals, yet humans demonstrate remarkably consistent perceptual and behavioral capabilities. This coexistence of individual neural variability with functional equivalence raises a fundamental question: to what extent do different brains employ shared computational principles when processing identical information? Understanding this relationship constitutes a central challenge in systems neuroscience, where identifying common neural motifs across individuals can reveal both universal coding mechanisms and the constraints that shape cortical organisation \citep{kriegeskorte2008representational,haxby2020hyperalignment}. Addressing cross-subject neural alignment requires methods that identify common representational structure while accommodating individual anatomical and functional differences to transform neural data from different individuals into a shared representational space enabling direct comparison \citep{haxby2011common,chen2015reduced}.

Emerging evidence suggests that such shared structure is a fundamental property of neural systems, not merely a convenient modelling assumption. \citet{safaie2023preserved} demonstrated that neural population latent dynamics are preserved across animals performing similar motor behaviours, arguing these reflect species-level constraints on circuit organisation. In visual cortex, hyperalignment recovers common high-dimensional representational spaces across subjects \citep{haxby2011common,haxby2020hyperalignment}, indicating that shared computational organisation persists despite idiosyncratic cortical topographies.

Cross-subject alignment enables population-level inference, identification of common neural motifs that reveal universal coding mechanisms, and characterisation of individual differences across subjects \citep{sucholutsky2023getting,safaie2023preserved,anderson2024hyperalignment,meshulam2021neural}. Beyond population analysis, aligned representations enable cross-subject neural prediction and universal decoding: neural responses from one subject can reconstruct another's brain activity, enabling decoders that generalise across individuals with minimal calibration \citep{dai2025mindaligner,scotti2024mindeye2, ferrante2024through, wang2024unibrain}, a capability particularly valuable for brain-computer interfaces and clinical applications where individual subjects cannot provide extensive training data.

Traditional alignment methods, e.g., Shared Response Models \citep[SRM;][]{chen2015reduced}, Procrustes analysis \citep{gower1975generalized}, and Hyperalignment \citep{haxby2011common}, derive transformations to perform alignment from neural responses to shared stimuli viewed by all subjects. While effective when substantial stimulus overlap exists, they face critical limitations: they require shared data for deriving transformations, and generalisation to non-shared stimuli depends entirely on representational coverage of the shared subset. For naturalistic paradigms where subjects view largely non-overlapping stimuli, these constraints become prohibitive and they preclude the large-scale, cross-study data aggregation that the increasing availability of open neuroimaging data makes both desirable and technically feasible. 

We introduce MED-VAE, a Multi-Encoder Multi-Decoder Variational Autoencoder achieving cross-subject alignment  of visual cortex responses to natural scenes in a common space \emph{without} requiring shared stimuli. Our key insight is that a pre-trained artificial neural network (ANN) processing identical images can serve as an alignment scaffold. By training subject-specific encoders to project fMRI responses into a shared latent space, and requiring this space to reconstruct ANN features via a shared decoder, we create implicit alignment pressure: all subjects' representations in the shared space must be compatible with the same computational structure encoding hierarchical visual features and semantic relationships. 
The contributions of our method are:
\begin{enumerate}
    \item \textbf{MED-VAE architecture:} it eliminates the need for shared stimuli between subjects through ANN-guided latent alignment.
    \item \textbf{Semantic organization:} our method is able to preserve behaviorally relevant categorical information in its aligned space, evidenced by higher silhouette scores and cross-subject category decoding accuracy.
    \item \textbf{Cross-subject alignment:} it achieves higher alignment and more robust generalization to held-out stimuli where traditional methods degrade.
    \item \textbf{Cross-trial reconstruction:} reconstructing from the aligned space back to each subject’s original neural space, it obtains equal cross-trial reconstruction revealing equal preservation of stimulus-related signal.
    \item \textbf{Cross-subject neural prediction and image decoding:} we show that we obtain more accurate cross-subject neura prediction and image decoding.
    \item \textbf{Framework for finding common task-relevant latent spaces :} 
     the shared (vision) latent space supports multiple downstream tasks, with potential for aggregating neural data across subjects and studies without requiring shared stimuli.
\end{enumerate}

\section{Related Work}
\label{sec:related}

\textbf{Traditional alignment methods.} Traditional alignment approaches transform individual subjects' representational spaces into unified frameworks where neural activity can be directly compared and analyzed 
to determine what information is systematically shared across individuals. Shared Response Model \citep[SRM;][]{chen2015reduced} decomposes neural responses into shared and subject-specific components through probabilistic matrix factorisation, learning transformation matrices from each subject's individual neural space to the common space, based on the responses to common stimuli. Procrustes analysis \citep{gower1975generalized} finds optimal orthogonal transformations minimising distances between corresponding points across subjects, enforcing rigid geometric transformations that preserve within-subject distances and angles. Hyperalignment \citep{haxby2011common} iteratively applies Procrustes across subjects to derive a common space. While effective for datasets with extensive stimulus overlap, these methods require substantial shared data for deriving alignment transformations, a strict constraint for naturalistic paradigms where individual subjects view largely non-overlapping stimuli. Moreover, generalisation to non-shared stimuli depends entirely on representational coverage of the shared subset; if alignment transformations learned from shared images fail to capture relevant neural dimensions, the remaining subject-specific responses remain suboptimally aligned.

\noindent\textbf{Pairwise cross-subject transfer methods.} Recent work has used pretrained ANNs to bypass the shared-stimulus requirement for pairwise cross-subject transfer. \citet{wang2025inter} use ANN features as a content-loss training signal for pairwise neural code converters, while \citet{wasserman2026functional} use per-subject ANN-based encoding models to synthesise paired fMRI data for training pairwise linear transformations. Crucially, both approaches create a separate transformation for each source--target subject pair, i.e., scaling as $O(n^2)$ for $n$ subjects, and neither constructs a shared representational space: the output is converted voxel patterns in the target subject's native brain space, not a common manifold in which all subjects can be simultaneously compared. These methods therefore address a complementary goal, i.e., pairwise brain-to-brain transfer, rather than the population-level aligned space that is the focus of this work. MED-VAE trains a single model across all subjects, producing a shared latent space supporting population-level analyses. Our empirical comparisons accordingly focus on methods that, like MED-VAE, construct common spaces, enabling evaluation on alignment metrics defined only within a shared representational framework.

\section{Methods}
\label{sec:methods}

Our approach rests on the finding that ANNs trained on visual tasks produce internal representations that predict neural responses in visual cortex \citep{kriegeskorte2015deep, yamins2016using, conwell2024large}. This model-brain correspondence means the ANN carries task-relevant structure making it a principled alignment scaffold. 
MED-VAE achieves cross-subject alignment implicitly through architectural constraints rather than explicit stimulus matching or prescribed correspondences. The framework constructs a common space wherein all subjects' latent representations must reconstruct, along with each subject's own neural activity via subject-specific decoders, a single shared ANN feature space via a common decoder. Simultaneously, the subject-specific fMRI decoders reconstruct their respective brain activity from this same unified ANN encoding, projected into the latent space via a shared ANN encoder (Fig. ~\ref{fig:medvae_archit}). This bidirectional constraint establishes pressure toward representational universality across all subjects without explicitly defined correspondence functions.

The ANN scaffold \textbf{overpasses the need for shared stimuli}, as each fMRI representation aligns to the ANN representation, despite not having identical correspondence to another subject. It also provides principled task-relevant structure, i.e., encoding categories, hierarchical features, and semantic relationships, that guides alignment toward computationally meaningful dimensions rather than arbitrary geometric correspondences. MED-VAE does not prescribe any relationship between ANN semantic structure and fMRI latent organisation; alignment emerges from reconstruction objectives.  Crucially, the ANN scaffold is required only during training to induce this alignment; at inference each subject is projected into the shared space by their fMRI encoder alone, without the ANN model. \color{black}

\subsection{MED-VAE Architecture}

\begin{figure*}[ht]
    \centering
    \includegraphics[width=0.82\textwidth]{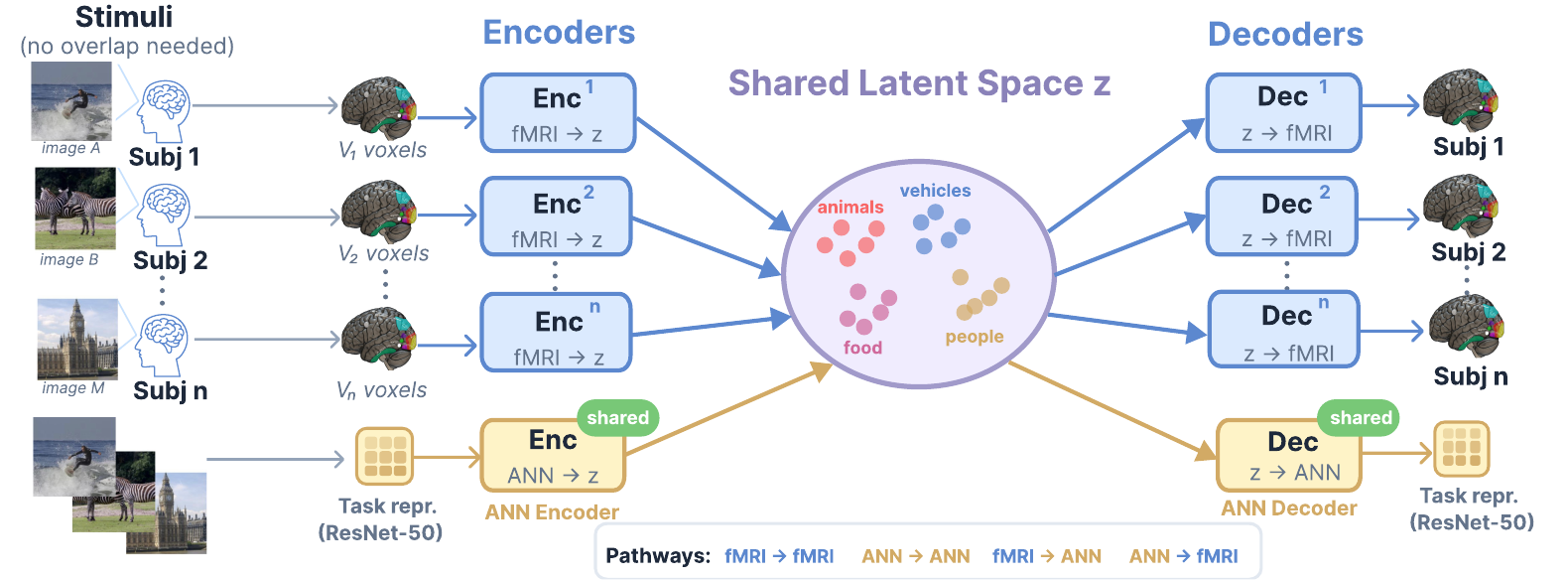}
    \caption{\textbf{Multi-Encoder Multi-Decoder Variational Autoencoder (MED-VAE)}. Subject-specific fMRI encoders map neural responses to a shared latent space; a shared ANN encoder maps ResNet-50 features to the same space. Subject-specific fMRI decoders reconstruct voxel patterns; a shared ANN decoder reconstructs ResNet-50 features from all latent representations, enforcing cross-subject alignment.}
    \label{fig:medvae_archit}
\end{figure*}

Our framework employs a Multi-Encoder Multi-Decoder Variational Autoencoder projecting subject-specific neural responses into a shared latent space organized by task-relevant representational structure. The architecture, as shown in Fig.~\ref{fig:medvae_archit}, includes four pathway 
\vspace{4pt} types:

\noindent \textbf{Subject-specific fMRI encoders.} For each subject $i \in \{1, \ldots, N\}$, an encoder $\text{Enc}_i^{\text{fMRI}}: \mathbb{R}^{V_i} \rightarrow \mathbb{R}^{d}$ maps fMRI responses to a shared $d$-dimensional latent space:
\begin{equation}
\mathbf{z}_i \sim \mathcal{N}(\boldsymbol{\mu}_i, \text{diag}(\boldsymbol{\sigma}_i^2)), \quad \boldsymbol{\mu}_i, \log \boldsymbol{\sigma}_i = \text{Enc}_i^{\text{fMRI}}(\mathbf{x}_i^{\text{fMRI}})
\end{equation}
\vspace{4pt}
where $V_i$ denotes voxel count for subject $i$.

\noindent \textbf{Shared ANN encoder.} A shared encoder $\text{Enc}^{\text{ANN}}: \mathbb{R}^{K} \rightarrow \mathbb{R}^{d}$ maps ResNet-50 features to the same latent space, establishing the computational structure toward which fMRI 
\vspace{4pt}
representations align.

\noindent \textbf{Subject-specific fMRI decoders.} Individual decoders $\text{Dec}_i^{\text{fMRI}}: \mathbb{R}^{d} \rightarrow \mathbb{R}^{V_i}$ reconstruct each subject's voxel responses from latent representations originating from both their respective fMRI 
\vspace{4pt}
encoder and the ANN encoder.

\noindent \textbf{Shared ANN decoder.} A universal decoder $\text{Dec}^{\text{ANN}}: \mathbb{R}^{d} \rightarrow \mathbb{R}^{K}$ reconstructs ANN 
features from any latent \vspace{4pt} representation.

The shared ANN decoder constitutes the \emph{primary alignment mechanism}: requiring all subjects' latent representations to reconstruct identical ANN features enforces cross-subject alignment within the latent space. For subjects $i$ and $j$ viewing not identical but similar stimuli, $\mathbf{z}_i$ and $\mathbf{z}_j$ must reconstruct similar ANN features through the shared decoder, enforcing $\mathbf{z}_i$ and $\mathbf{z}_j$ to be close in the latent space, implicitly resulting in cross-subject alignment. A \emph{secondary alignment mechanism} arises from the subject-specific fMRI decoders needing to reconstruct neural activity from both same-subject fMRI and ANN-encoder latent representations. The shared ANN encoder output serves as a universal input that regularizes subject-specific fMRI decoders toward functionally equivalent transformations, making them more compatible with other subjects' fMRI encoders, which is beneficial for cross-subject neural prediction.

Notably, at inference the ANN scaffold is discarded: a subject's responses are mapped into the shared space by their fMRI encoder alone $\text{Enc}_i^{\text{fMRI}}(\mathbf{x}_i^{\text{fMRI}})$, and all the inference use cases we demonstrate below operate without requiring the ANN model. The ANN-derived scaffold is only used to encourage latent alignment during training. \color{black}
Another thing to note is that because encoders and decoders are subject-specific with independently parameterised input/output dimensions, the architecture naturally accommodates subjects with different voxel counts or ROI definitions, including, in principle, subjects from different studies or scanners.

\setlength{\textfloatsep}{8pt} 

\subsection{Training Objective}

The MED-VAE loss integrates the four reconstruction pathways with KL regularization:
\begin{align}
\mathcal{L} &=
\underbrace{\mathcal{L}_{\text{fMRI}\rightarrow\text{fMRI}}}_{\text{Within-modality fMRI}}
\mkern-4mu+\mkern-4mu
\underbrace{\mathcal{L}_{\text{ANN}\rightarrow\text{fMRI}}}_{\text{Cross-modal: ANN\textrightarrow fMRI}}
\label{eq:loss} \\
&\quad
\mkern-4mu+\mkern-4mu
w \Big[
\underbrace{\mathcal{L}_{\text{ANN}\rightarrow\text{ANN}}}_{\text{Within-modality ANN}}
\mkern-4mu+\mkern-4mu
\underbrace{\mathcal{L}_{\text{fMRI}\rightarrow\text{ANN}}}_{\text{Cross-modal: fMRI\textrightarrow ANN}}
\Big]
\mkern-4mu+\mkern-4mu
\mathcal{L}_{\text{KL}}
\nonumber
\end{align}

The weighting parameter $w$ controls alignment pressure through the shared ANN decoder pathway. Higher $w$ values strengthen the requirement that fMRI encoders produce latent codes compatible with the shared ANN decoder. All results reported here use $w = 5$,  selected on a validation set as the value maximising downstream alignment metrics while maintaining NC-normalised reconstruction above 100\% (Appendix~\ref{app:w_sweep}). Detailed training objective equations can be found in Appendix \ref{appendix:training_obj}. 

\subsection{Evaluation Metrics}

\textbf{Alignment quality.} Component-wise correlation measures dimension-specific alignment across subjects and Representational Similarity Analysis (RSA) evaluates geometric structure similarity between subjects' latent representations. Both metrics are computed for subject pairs in the common aligned space. More information on the metrics can be found in Appendix \ref{Appendix:Metrics}.
\vspace{3pt}

 \noindent \textbf{Within-subject reconstruction.} Within-subject reconstruction is evaluated by calculating the correlation between the original subject data (input) and the reconstructed subject data (output), i.e., when projecting from the common space back to each subject's original neural space, thus quantifying the level of subject-specific neural information preserved in the common space. This correlation can be evaluated within-trial or cross-trial, i.e., correlate the reconstructed output with the original data of the same trial (or trial averages of the same stimuli) or a different trial for the same stimulus.
\vspace{3pt}

 \noindent \textbf{Cross-trial evaluation.} Within-trial metrics 
 allow noise artifacts, e.g., thermal noise, moment-to-moment physiological fluctuations (respiration phase, cardiac phase), and cognitive state variations, that are in both the input and the reconstructed output, to inflate performance by reconstructing this type of noise in addition to stimulus-related signal. Cross-trial evaluation addresses this confound by correlating reconstructions from Trial $i$ with independent Trial $j \neq i$ of identical stimuli, breaking the correlation between input noise and evaluation target.
 \vspace{3pt}

\noindent \textbf{NC-normalised reconstruction.} To quantify whether a method's reconstruction reflects stimulus-driven signal or trial-specific noise, we normalise each voxel's reconstruction correlation by its per-voxel correlation noise ceiling derived from the NSD-provided ncsnr values. Values above 100\% indicate reconstruction of non-stimulus-driven variance. Full derivation in Appendix~\ref{app:nc_norm}. 

\vspace{3pt}

 \noindent \textbf{Category encoding.} We evaluate semantic organisation in the common space using two complementary metrics. Silhouette scores, adapted for multi-label images (Appendix~\ref{Appendix:Metrics}), are computed on the pooled latent representations of all subjects, quantifying how coherently stimuli cluster by semantic category in the shared space. Additionally, leave-one-subject-out multi-label logistic regression decoding, reporting exact match and balanced accuracy, assesses whether category-discriminative structure generalises across subjects.
\vspace{4pt}

 \noindent \textbf{Cross-subject neural prediction.} For subject pairs $(i,j)$, we compute $\text{corr}(\text{Dec}_j^{\text{fMRI}}(\mathbf{z}_i), \mathbf{x}_j^{\text{fMRI}})$: how accurately subject $i$'s latent representation, decoded through subject $j$'s decoder, reconstructs subject $j$'s actual neural responses.

\subsection{Statistical Analysis}
\label{section:stat}
For all metrics we report $p$-values and effect sizes (Cohen's $d_z$).
Methods were compared with two-tailed paired $t$-tests over the four subjects ($n=4$). Reconstruction and decoding metrics are naturally per-subject; component correlation and RSA are pairwise, but as these values are
not independent, e.g., in pairwise RSA each subject appears in three pairs, the tests use one value per subject, i.e, its mean
correlation with the other three subjects. 
\color{black}
\section{Experimental Setup}
\label{sec:setup}

\paragraph{fMRI Dataset.} We use the Natural Scenes Dataset \citep[NSD;][]{allen2022massive}, comprising 8 subjects viewing mostly unshared $\sim$10,000 natural images each across 30--40 scanning sessions, with $\sim$1,000 images shared among all or a subset of subjects. The NSD data were downloaded from MIT Algonauts Project 2023 Challenge \citep{gifford2023algonauts}.

Cross-subject evaluation focuses on two image sets: 872 images viewed by all participants (shared) and 128 images viewed by a subset of four participants, subjects 1, 2, 5 and 7 (semi-shared). 
MED-VAE is trained exclusively on subject-specific 
(non-overlapping) responses.
MED-VAE therefore never 
observes cross-subject correspondences during training, i.e., it 
cannot exploit stimulus overlap to learn alignment, relying 
entirely on the ANN scaffold to discover shared structure. 
SRM and Procrustes, by contrast, derive their alignment 
transformations from the 872 shared images.
The held-out 128 images can be used as a generalization test for all methods and a proxy for the cross-subject alignment quality of the non-shared images.

For ROI selection, we followed \citet{conwell2024large}, where neural responses derive from occipitotemporal cortex ($\sim$20,730 voxels per subject) in fsaverage surface space, encompassing object- and category-selective regions (LOC, FFA, PPA, EBA).
For the cross-subject image decoding evaluation, we use a frozen MindEye2 decoder \citep{scotti2024mindeye2}; for this evaluation we use the ROIs specified in the original MindEye2 work rather than our standard ROIs.

\textbf{ANN features.} ResNet-50 activations (pretrained on ImageNet)  were extracted for all the images viewed by subjects. We chose ResNet-50 as it has  been shown to produce representations particularly well-aligned with human high-level visual cortex \citep{conwell2024large}. We sampled one layer every 8 to reduce dimensionality of data while obtaining activations from all depths. Each selected layer's activations were reduced via Sparse Random Projection (SRP) to 3,500 dimensions each.
Core alignment metrics vary <2\% across SRP seeds, confirming SRP seed choice is not a meaningful variance source. We assesed the robustness of MED-VAE framework on the use of different ANNs as scaffold, by training with CLIP ViT-L/14 (contrastive Vision Transformer). This analysis can be found on Appendix ~\ref{app:scaffold}.  

\textbf{Model Training.} 
 Encoders and decoders are two-layer MLPs with ReLU activations and dropout ($p = 0.3$). At $d_{\text{latent}} = 32$, the hidden dimension is $d_{\text{hidden}} = 256$, giving layer widths of $input\, dimension \rightarrow 512 \rightarrow 256 \rightarrow 32$ (encoder) and $32 \rightarrow 256 \rightarrow 512 \rightarrow output\,dimension$ (decoder). 
In experiments where $d_{\text{latent}} = 512$ (Section~5.5), $d_{\text{hidden}}$ scales to $1024$. Models were trained for 30 epochs with Adam ($\text{lr} = 10^{-4}$, batch size $128$).
Full architectural details are given in Appendix~\ref{appendix:model_arch}.

 \textbf{Classical alignment baselines.} We compared MED-VAE against Shared Response Model (SRM; \citet{chen2015reduced}) and generalized Procrustes analysis  \citep{gower1975generalized, haxby2020hyperalignment}, each learning a per-subject linear map into a common $d_{\text{latent}} = 32$ space. Both were fit on the 872 images viewed by all subjects (shared) and evaluated on the 128-image set shared only by subjects {1,2,5,7} (semi-shared); MED-VAE, by contrast, is trained on each subject's full non-shared responses, the data volume the baselines cannot exploit by construction. Because Procrustes is dimensionality-preserving, i.e., its output dimension equals its input dimension, we first reduced each subject to the $d_{\text{latent}} = 32$ latent with PCA, then aligned subjects iteratively, namely rotating each to the current group-mean configuration via the orthogonal-Procrustes solution and re-estimating the mean until convergence. SRM is itself dimensionality-reducing, and we used its deterministic variant (DetSRM).\color{black}
\section{Results}
\label{sec:results}

\begin{figure*}[!t]
    \centering
    \includegraphics[width=\textwidth]{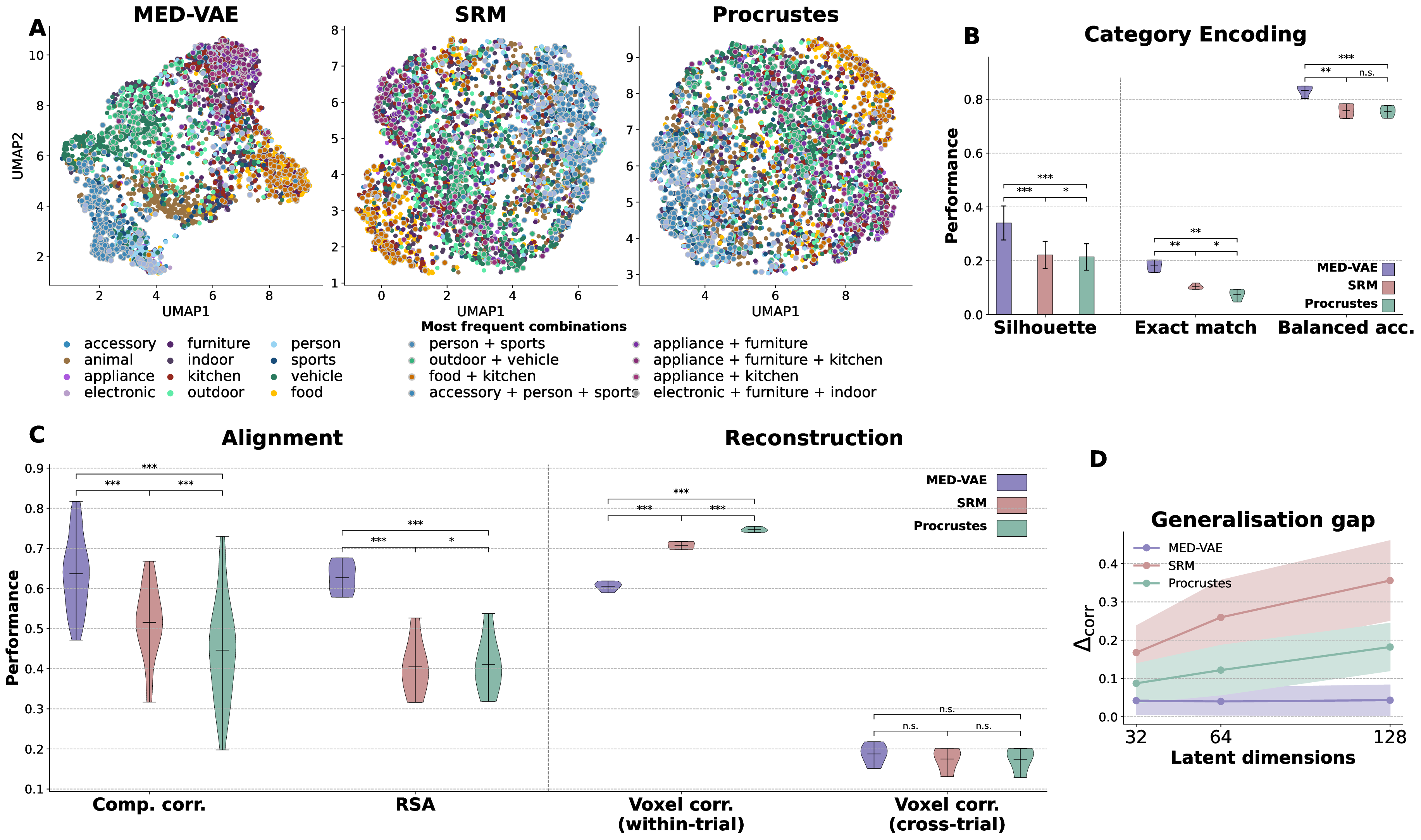}
    \caption{\textbf{MED-VAE yields a more semantically structured latent manifold, facilitating superior cross-subject alignment and enhanced generalization to unseen stimuli.}
    \textbf{A.} Common subspaces found by MED-VAE, SRM and Procrustes. 
    MED-VAE latent space better encodes stimulus category information. 
    Multi-label images plotted as multiple jittered dots coloured by label; 
    the eight most frequent label combinations shown as single grey-edged dots 
    with blended colours. 
    \textbf{B.} Category encoding quality in the shared space. 
    Silhouette scores measuring category cluster separability in the pooled multi-subject latent space;  bars show the bootstrap mean with 95\% CI, and the silhouette significance bracket is a paired image-bootstrap over the 128 images (B = 5000; two-sided $p$). Exact match and balanced accuracy for leave-one-subject-out multi-label category decoding; violins show the per-subject values and their brackets denote two-tailed paired t-tests across subjects $(n = 4)$. \color{black} MED-VAE achieves superior category 
    encoding across all metrics.
    \textbf{C.} Performance comparison across alignment and reconstruction 
    metrics. \textit{Alignment:} component-wise correlation and RSA. 
    \textit{Reconstruction:} within-trial and cross-trial voxel 
    correlation---traditional methods' within-trial advantage disappears 
    under cross-trial evaluation, indicating the reconstruction of within-trial noise rather than stimulus-related signal.
    Significance brackets denote two-tailed paired $t$-tests across subjects ($n=4$).  For the
    alignment metrics the violins show each metric's native distribution---the 32 component-wise
    correlations (component correlation) and the 6 subject-pair values (RSA)---but, because these
    component- and pair-level values are not independent observations, e.g., in RSA each subject appears in three
    pairs, the tests are run on one value per subject, e.g., each subject's mean correlation with the other
    three ($n=4$); reconstruction metrics are intrinsically per-subject ($n=4$). Stars reflect
    uncorrected $p$ ($^{***}p<0.001$, $^{**}p<0.01$, $^{*}p<0.05$; n.s., not significant).
    \color{black}
    \textbf{D.} Generalisation gap: difference in component-wise correlation 
    between images used for alignment training for SRM/Procrustes and 
    held-out images, across latent dimensionalities (32, 64, 128). 
    Shaded regions show $\pm 1$ s.d.\ across components. SRM and Procrustes 
    exhibit pronounced generalisation degradation on unseen stimuli. MED-VAE has not used any of these subsets for alignment training.}
    \label{fig:combined}
\end{figure*}

\subsection{Enhanced Semantic Organisation in Aligned Spaces}
\label{sec:categorical}
We first examine whether aligned common spaces preserve behaviourally relevant categorical structure. Fig.~\ref{fig:combined}A shows fMRI projections onto latent spaces produced by MED-VAE, SRM, and Procrustes, coloured by semantic category. MED-VAE latent spaces exhibit better categorical organisation: semantically related stimuli cluster coherently, with clear separation between semantic clusters. In contrast, SRM and Procrustes produce diffuse representations with substantial category overlap and poorly defined cluster boundaries.

To quantify the degree of semantic structure, we used two complementary metrics. Silhouette scores are computed on the pooled latent representations of all subjects, measuring how coherently stimuli cluster by semantic category in the shared space, (Fig.~\ref{fig:combined}B): 
MED-VAE achieves a markedly higher silhouette than both SRM and Procrustes (0.34 [0.29, 0.42] vs 0.22 [0.18, 0.28] and 0.21 [0.18, 0.27]; non-overlapping bootstrap 95\% CIs, paired image bootstrap B=5000, p<0.001 for both).
Additionally, leave-one-subject-out decoding assesses whether category-discriminative structure generalises across individuals: a multi-label classifier is trained on three subjects' aligned representations and tested on the held-out subject's. 
MED-VAE yields higher exact-match accuracy than SRM and Procrustes (0.184 vs.\ 0.104 and 0.074, respectively; paired $t$-tests: $d_z = 4.6$ and 3.3, $p = 0.003$ and 0.007) and higher balanced accuracy (0.833 vs.\ 0.758 and 0.754, respectively; $d_z = 6.3$ and 10.5, $p = 0.001$ and 0.0002). 
To demonstrate that these results require task-relevant representations, we trained a MED-VAE framework with the activations of an \emph{untrained ResNet}; we observe no cross-subject category clustering in the latent space; 
(Fig.\ref{fig:untrained}, Appendix~\ref{app:control}). As an additional control, we trained MED-VAE after shuffling the 1:1 correspondence between ResNet and fMRI data. We observe that representations cluster by subject identity rather than semantic content (Fig.\ref{fig:untrained}, Appendix~\ref{app:control}). In both control cases, silhouette score when combining all subjects' representations is near zero.

This enhanced semantic organization emerges from the ANN scaffold: the shared decoder requires neural latent representations to reconstruct ResNet-50 features encoding hierarchical visual structure and categorical boundaries. VAEs inherently produce latent spaces where semantically similar inputs cluster together, a property that holds even for ANN-to-ANN or image-to-image autoencoders. MED-VAE leverages this property while allowing the bidirectional pathways, i.e, fMRI\textrightarrow ANN, ANN\textrightarrow fMRI, fMRI\textrightarrow fMRI, to jointly shape the latent geometry.
These results demonstrate that a common representational space exists across subjects preserving the computational organisation of visual representation, i.e., categorical structure and semantic relationships, not merely arbitrary statistical correspondence. The ANN-informed scaffold is essential: as our untrained ResNet control confirms, this organisation depends on the scaffold carrying meaningful computational structure, not on the VAE or ResNet architecture alone.

\subsection{Cross-Subject Alignment with Robust Generalisation}

MED-VAE achieves higher cross-subject alignment than SRM and Procrustes, respectively, as measured by both component-wise correlation (0.64 vs.\ 0.52 and 0.45; paired $t$-tests: $d_z = 7.7$ and 11.4, $p = 0.0006$ and 0.0002) and RSA (0.63 vs.\ 0.41 and 0.41; $d_z = 9.0$ and 8.1, $p = 0.0004$ and 0.0005); Fig.~\ref{fig:combined}C, Alignment panel.
$t$-tests were performed on metrics computed per subject ($n = 4$), as described in Section \ref{section:stat}.
\color{black}
Critically, MED-VAE maintains superior performance despite being trained purely with non-overlapping stimuli across subjects. To quantify this effect and contrast it with standard benchmarks, we calculated a \textit{generalization gap}, defined as the difference in component correlation between the shared ($N_{shared} = 872$) and semi-shared ($N_{semi-shared} = 128$) image sets: $\Delta_{\text{corr}} = \rho_{\text{shared}} - \rho_{\text{semi-shared}}$. While both SRM and Procrustes are fitted directly on the 872 shared images, they exhibit weak generalization that scales poorly with latent dimensionality $d$ (Fig.~\ref{fig:combined}D). In contrast, MED-VAE demonstrates robust generalization, indicating that it captures subject-invariant representations rather than over-fitting to stimulus-specific features.

This generalization gap exposes an important flaw in classical alignment: by deriving transformations solely from shared stimuli, methods like SRM and Procrustes fail to capture the latent dimensions required to align the broader dataset. Without ground-truth correspondences for non-shared stimuli, the 128-image test set serves as a vital proxy for global alignment quality.
In contrast, MED-VAE leverages an ANN scaffold to extract alignment pressure from the entire stimulus set, eliminating the risk of overfitting to a "privileged" shared subset. By imposing computational structures—such as categorical hierarchies and representational geometries—the scaffold forces subject convergence into a unified manifold.  Per-layer analysis (Appendix~\ref{app:per-layer}) confirms that brain-ANN alignment increases monotonically with scaffold depth, with late layers, which encode categorical and object-selective representations, reaching 80\%. 

\subsubsection{Cross-Subject Latent Space Retrieval}
Cross-subject retrieval in the shared latent space provides a direct test of whether aligned representations preserve stimulus identity across individuals. For each image in the 128 semi-shared image dataset, we used subject A's latent representation as a query and retrieved the nearest neighbour among subject B's latent representations, measuring whether the retrieved embedding corresponded to the same image.
MED-VAE achieved substantially higher retrieval accuracy than both SRM and Procrustes across all top-$k$ thresholds, i.e., more than double the top-1 accuracy of either baseline (46.2\% vs 21.8\% and 18.4\%), as shown in Table~\ref{tab:retrieval}.
 \color{black}
This reflects both coherent categorical structure, as also shown in Section~\ref{sec:categorical}, and stimulus-level alignment enforced by the ANN scaffold.

\begin{table}[t]
\centering
\caption{Average cross-subject retrieval top-k accuracy on the 128 semi-shared held-out dataset. MED-VAE substantially outperforms traditional alignment methods at preserving stimulus identity across subjects.}
\label{tab:retrieval}
\begin{tabular}{lcccc}
\toprule
Method & Top-1 & Top-2 & Top-5 & Top-10 \\
\midrule
Chance & 0.8\% & 1.6\% & 3.9\% & 7.8\% \\
Procrustes & 18.4\% & 25.6\% & 40.9\% & 54.1\% \\
SRM & 21.8\% & 30.0\% & 44.5\% & 57.8\% \\
\textbf{MED-VAE} & \textbf{46.2\%} & \textbf{62.0\%} & \textbf{80.3\%} & \textbf{91.1\%} \\
\bottomrule
\end{tabular}
\end{table}
\subsection{Cross-Trial Evaluation Reveals Preservation of Signal}

A key question is whether MED-VAE's more aligned, semantically better-organized common space preserves the same level of stimulus-driven neural information. Within-trial reconstruction, where models reconstruct from trial-averaged input and are evaluated against the same trial average, suggests a trade-off:
SRM and Procrustes achieve significantly higher voxel correlations than MED-VAE (0.71 and 0.75 vs.\ 0.61, respectively; paired $t$-tests: $|d_z| = 17.7$ and 21.0, both $p < 0.001$, $n = 4$ subjects; Fig.~\ref{fig:combined}C, Reconstruction panel).
\color{black} 
However, cross-trial evaluation, which isolates stimulus-driven signal by correlating reconstructions from one trial with ground truth from independent trials of the same stimulus (thereby removing noise correlations between input and evaluation target), reveals no significant differences between methods (0.19/0.18/0.17; all $p > 0.11$)\color{black}. The expected level difference between within- and cross-trial values given these ROIs' noise ceiling SNR is calculated in Appendix~\ref{app:cross_trial_ceiling} and agrees with the observed gap.
The difference reflects the noise component that within-trial metrics erroneously credit as signal, while also the fact that independent realizations of noise can be negatively correlated. 
This cross-trial result demonstrates that traditional methods' within-trial advantage derives from averaged trial-specific measurement noise rather than stimulus-related signal. 

NC-normalised self-reconstruction provides a complementary perspective on the same finding, i.e., normalising each voxel's reconstruction correlation by its per-voxel correlation noise ceiling (derived from the NSD-provided noise ceiling SNR; see Appendix~\ref{app:nc_norm} for the full formulation), with values above
100\% indicating reconstruction of non-stimulus-driven variance. Procrustes and SRM achieve 154.8\%
 and 
138.3\% respectively, substantially exceeding the ceiling, while MED-VAE achieves 102.5\%, confirming that the within-trial gap is attributable to noise reconstruction rather than superior signal preservation. 
MED-VAE, thus, does not sacrifice signal for alignment; it suppresses noise while preserving stimulus-driven neural structure. MED-VAE achieves aligned spaces with superior semantic organization, superior alignment, and comparable preservation of stimulus-driven signal.

\subsection{Higher Alignment Enables Superior Cross-Subject Neural Prediction}

\begin{figure}[ht]
    \centering
    \includegraphics[width=\columnwidth]{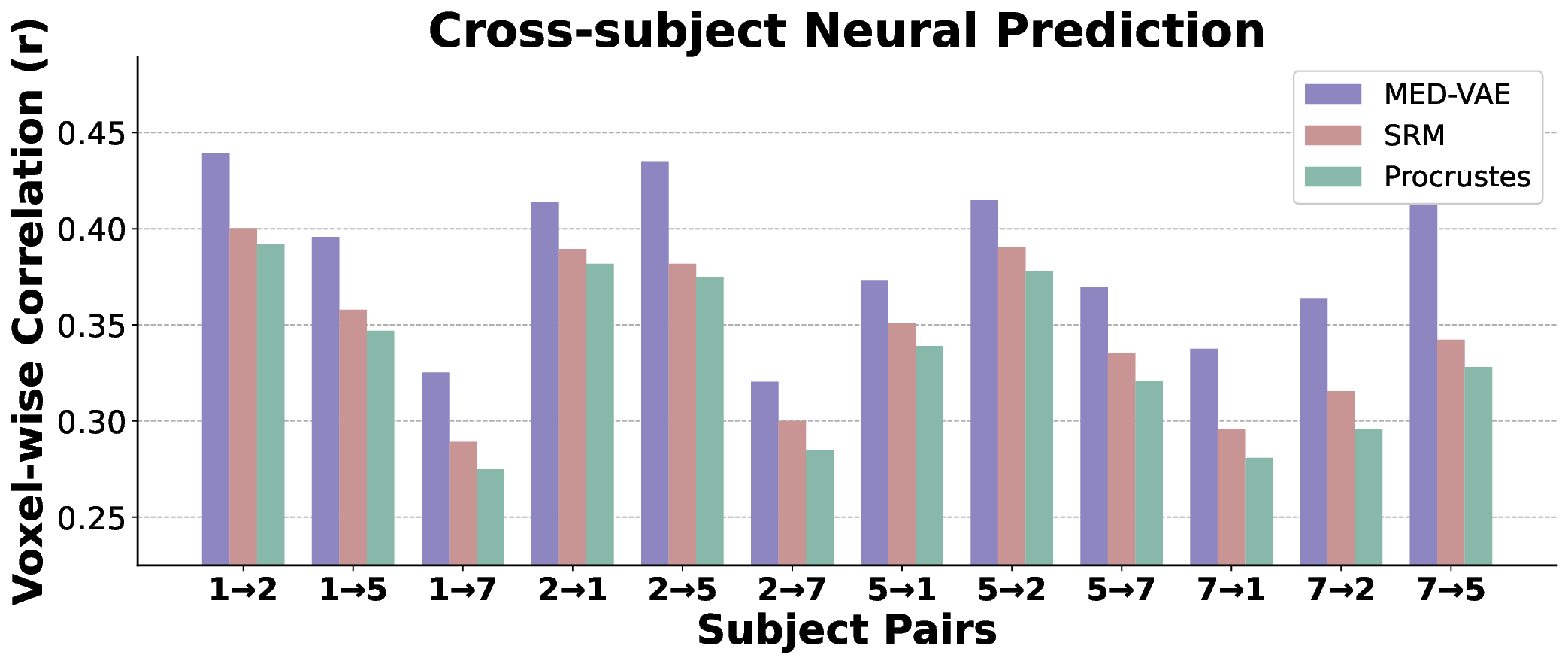}
    \caption{\textbf{Cross-subject functional alignment comparison via fMRI Voxel Correlation}. Voxel-level prediction accuracy across twelve subject transfer pairs ($i \rightarrow j$). MED-VAE (purple) outperforms SRM and Procrustes. Results are on the 128 held-out images.}
    \label{fig:cross_subj_barplot}
\end{figure}

Cross-subject neural prediction, i.e., using subject $i$'s latent representation with subject $j$'s decoder to reconstruct subject $j$'s voxel-level responses, provides functional validation of alignment quality. This metric directly tests whether the common space captures genuinely shared neural structure generalisable across individuals. MED-VAE outperforms the other methods as shown in Fig.~\ref{fig:cross_subj_barplot} for all pairs of subjects. This performance advantage follows directly from the properties established above: a semantically organised, well-aligned common space that preserves stimulus-driven signal naturally supports accurate translation of neural representations between individuals.

\setlength{\textfloatsep}{10pt} 

\begin{table*}[h]
\caption{Average cross-subject image decoding performance across all source--target pairs (subjects 1, 2, 5, 7). 
Higher is better ($\uparrow$) for all metrics except Eff and SwAV ($\downarrow$).
}
\label{tab:avg_norefine}
\centering
\begin{tabular}{lcccccccccc}
\toprule
Method
& PixCorr $\uparrow$
& SSIM $\uparrow$
& Alex2 $\uparrow$
& Alex5 $\uparrow$
& Incep $\uparrow$
& CLIP $\uparrow$
& Eff $\downarrow$
& SwAV $\downarrow$
& Fwd\% $\uparrow$
& Bwd\% $\uparrow$ \\
\midrule
VAE 32D
& \textbf{0.160}
& 0.338
& \textbf{0.846}
& \textbf{0.925}
& \textbf{0.870}
& \textbf{0.827}
& \textbf{0.776}
& \textbf{0.463}
& \textbf{76.7}
& \textbf{64.4} \\
SRM 32D
& 0.113
& 0.342
& 0.774
& 0.862
& 0.769
& 0.731
& 0.868
& 0.529
& 64.3
& 55.9 \\
Proc 32D
& 0.107
& \textbf{0.345}
& 0.758
& 0.836
& 0.736
& 0.699
& 0.885
& 0.541
& 49.5
& 45.3 \\
\midrule
VAE 512D
& \textbf{0.166}
& 0.344
& \textbf{0.867}
& \textbf{0.946}
& \textbf{0.904}
& \textbf{0.856}
& \textbf{0.742}
& \textbf{0.440}
& \textbf{86.5}
& 76.3 \\
SRM 512D
& 0.156
& 0.355
& 0.859
& 0.928
& 0.835
& 0.800
& 0.806
& 0.485
& 83.1
& \textbf{81.1} \\
Proc 512D
& 0.152
& \textbf{0.355}
& 0.845
& 0.916
& 0.825
& 0.786
& 0.816
& 0.489
& 81.3
& 80.3 \\
\bottomrule
\end{tabular}
\end{table*}

\subsection{Cross-subject Image Decoding as a Semantic Probe}

Cross-subject neural prediction evaluates alignment quality at the voxel level but does not assess what kind of information the predicted responses carry. We therefore used cross-subject image decoding to probe whether cross-subject predictions preserve sufficient semantic content to drive a complex downstream task.   The output of cross-subject neural prediction, i.e., subject
j's voxel-level responses reconstructed from subject
i's latent representation via subject
j's decoder (Section 5.4), is fed into a frozen MindEye2 decoder \citep{scotti2024mindeye2}  pre-trained exclusively on subject
j's data.   No MindEye2 parameters are updated at any stage.   
Neither MED-VAE nor SRM or Procrustes include decoding-related objectives during training. Image decoding is used here purely as a probe of what semantic information survives the cross-subject prediction pipeline for each method, not as a benchmark against purpose-built decoding systems.

\begin{figure}[ht]
    
    \centering   \includegraphics[width=\columnwidth]{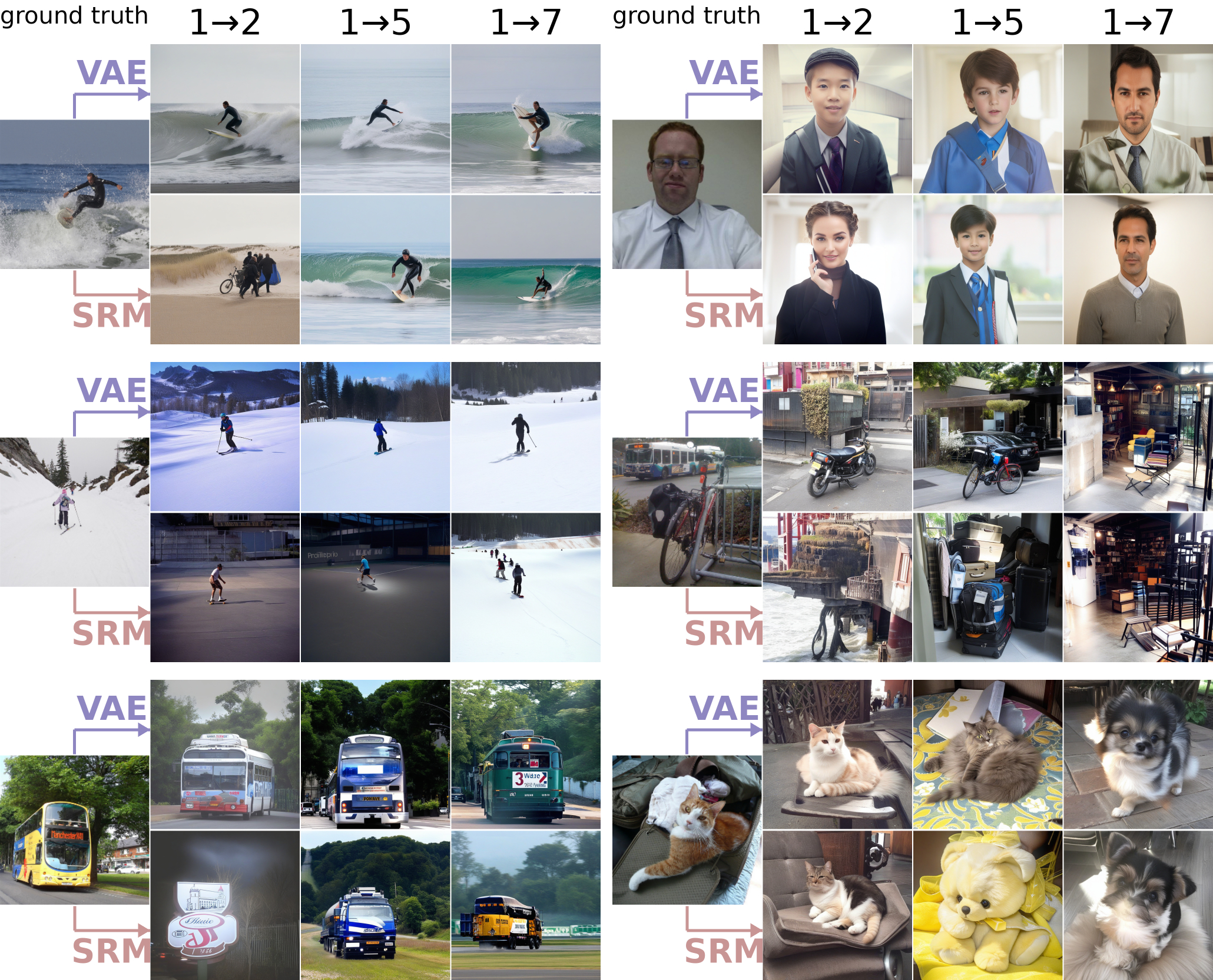}
    \caption{\textbf{Cross-subject image reconstruction examples comparing MED-VAE and SRM alignment}. For six selected stimuli, reconstructions are shown for three transfer pairs where subject 1 serves as the source (1→2, 1→5, 1→7). Ground truth images are displayed in the leftmost column. For each stimulus, the top row shows MED-VAE-based reconstructions and the bottom row shows SRM-based reconstructions.  Results are for the 512D setting. }
    \label{fig:reconstr_main}
\end{figure}

We evaluated all pairwise transfer configurations among subjects 1, 2, 5, and 7 on held-out images, using the NSD ROIs specified in the original MindEye2 work rather than our standard occipitotemporal ROIs, to ensure compatibility with the pre-trained decoder weights.
Table~\ref{tab:avg_norefine} presents results at 32D and 512D for the latent dimension, i.e., the dimension of the subjects' aligned space. At 32D, MED-VAE substantially outperforms both baselines across nearly all metrics, with the advantage most pronounced on high-level semantic measures and retrieval accuracy Fwd\%. At 512D, the gap between methods narrows considerably though MED-VAE retains a clear advantage on high-level measures (CLIP: 85.6\% vs 80.0\% vs 78.6\%; Inception: 90.4\% vs 83.5\% vs 82.5\%).
This dimensionality-dependent pattern reflects MED-VAE's semantic scaffold: at 32D, it already carries semantically relevant information, while SRM/Procrustes require higher capacity to separate signal from noise. That MED-VAE at 32D surpasses both baselines at 512D on high-level metrics underscores the efficiency of scaffold-guided alignment.
The baselines' relative strengths are confined to SSIM, a metric with known limitations, as \citet{scotti2024mindeye2} showed even degenerate reconstructions can score well on it, and backward retrieval at 512D, which rewards pattern distinctiveness rather than semantic fidelity. The detailed results for each pair of subjects and information on the metrics can be found in Appendix~\ref{Appendix:image_decoding}.
Qualitative examples of cross-subject image reconstructions are shown in Fig.\ref{fig:reconstr_main} for selected transfer pairs and with the full set of all 12 transfer directions provided in Fig.~\ref{fig:reconstructed_images} in Appendix~\ref{fig:reconstr_appendix}. MED-VAE reconstructions more consistently preserve the semantic content of the original stimuli and exhibit smaller variance across different transfer pairs.

\section{Discussion and conclusions}
\label{sec:discussion}

\paragraph{Population-level neuroscience implications:} Aligning subjects without shared stimuli into a semantically organised common space enables population-level analyses that were previously impractical. Data from subjects scanned under different protocols, viewing different stimuli, could in principle be projected into a common representational framework, enabling group-level analyses of representational geometry and individual differences analysis at the neural pattern level rather than the summary-statistic level. 
 MED-VAE provides response-level alignment without shared stimuli, and the resulting common space exhibits qualitatively superior semantic organisation. While the current experiments use subjects from the same dataset, the architecture naturally accommodates heterogeneous inputs, e.g.,  different voxel counts, ROI definitions, and in principle, different scanners, making cross-study alignment a natural next step.
 \vspace{-5ex}
 \paragraph{Cross-subject neural prediction and downstream applications:} Superior alignment translates to more accurate prediction of individual subjects' neural responses from other subjects' representations. This enables practical applications including cross-subject image decoding with minimal per-subject calibration, data augmentation for subjects with limited scanning time by borrowing aligned data from other subjects, and clinical applications where scanning time is constrained.  A new participant can be integrated into an existing shared space by training only a subject-specific encoder–decoder pair on their own non-overlapping stimuli recovering $\sim$ 98\% of the jointly-trained alignment quality, with strong alignment retained from as little as 10\% of the NSD experiment scanning time without retraining the shared space. More information on this analysis can be found in Appendix \ref{app:novel-subjects}.
 
  \vspace{-2.5ex}
 \paragraph{Potential extensions  beyond vision:} The ANN scaffold approach  could, in principle, be extended to  other domains where good model-brain alignment exists,  though this remains to be demonstrated. Encoding models for language processing \citep{antonello2023scaling, caucheteux2022brains}, auditory processing \citep{kell2018task}, and multimodal cognition \citep{tang2023brain} demonstrate that suitable scaffolds exist across cognitive domains. 
 Extension to temporal domains such as language and auditory processing would require adapting the framework from trial-level static responses to continuous temporal streams, a non-trivial but tractable modification.
  \vspace{-3ex}
 \paragraph{Limitations and future work:} Several limitations warrant discussion. All subjects in the current study are drawn from NSD, sharing scanner hardware and preprocessing pipelines; cross-study alignment with genuinely heterogeneous acquisition parameters remains to be demonstrated, though the architecture is designed to support it. The ResNet-50 scaffold imposes a particular computational lens on alignment: the common space can only capture shared neural structure that is reflected in the scaffold's representational geometry. Consequently, the quality of cross-subject alignment is fundamentally bounded by the scaffold's ability to explain neural responses. Using scaffolds with stronger model-brain correspondence, e.g., recurrent architectures that better capture the dynamics of ventral-stream processing \citep{kietzmann2019recurrence}, represents a natural direction for improving alignment quality. Systematic comparison across scaffold architectures would clarify which computational features are most relevant for cross-subject alignment. Finally, behavioral analysis in the latent space would strengthen the claim that the common space captures not just representational content but processing dynamics with behavioral relevance.
 \vspace{-3ex}
 \paragraph{Conclusions:} MED-VAE bypasses the shared-stimulus bottleneck of traditional neural alignment by replacing explicit matching with implicit constraints from a pretrained ANN scaffold. This architecture yields a cascade of advantages: superior semantic organization, robust generalization to held-out stimuli where classical methods fail, and preserved stimulus-driven signal fidelity. Overall, our framework offers an approach to identify task-informed manifolds shared across human brains that can be used for downstream tasks and neuroscientific investigation.

\noindent \paragraph{Reproducibility:}  The code is available at \url{https://github.com/pangelu9/MEDVAE_NSD}.
\section*{Acknowledgments}

A.P. is supported by the EPSRC Centre for Doctoral Training in Health Data Science [EP/S02428X/1]. R.P.C is funded by a ERC-UKRI Frontier Research Guarantee Starting Grant (EP/Y027841/1).

\printbibliography

\appendix

\section{Metrics}
\label{Appendix:Metrics}
 
\paragraph{Component-wise Correlation.}
To quantify the consistency of individual latent dimensions across subjects,
we computed component-wise Pearson correlation. For each latent dimension
$d \in \{1, \dots, D\}$, we extracted the activation vector
$\mathbf{z}^{(s)}_d \in \mathbb{R}^{N}$ across all $N$ test images for each
subject $s$. We then computed the pairwise Pearson correlation coefficient
$r(\mathbf{z}^{(s_i)}_d, \mathbf{z}^{(s_j)}_d)$ for all subject pairs
$(s_i, s_j)$ and averaged across pairs to obtain a single correlation value
per component. This metric assesses whether each latent dimension encodes
consistent information across subjects, independent of the remaining
dimensions.

\paragraph{Representational Similarity Analysis (RSA).}
We used RSA~\citep{kriegeskorte2008representational} to compare the
representational geometry across subjects. For each subject $s$, we
constructed a representational dissimilarity matrix (RDM)
$\mathbf{D}^{(s)} \in \mathbb{R}^{N \times N}$, across all N test
images, where each entry
$D^{(s)}_{ij} = 1 - r(\mathbf{z}^{(s)}_i, \mathbf{z}^{(s)}_j)$
corresponds to the Pearson correlation distance between the latent
representations of fMRI responses to images $i$ and $j$. Pairwise RDM similarity between
subjects was then quantified using the Pearson correlation between the
upper-triangular elements of their respective RDMs.

\paragraph{Clustering Quality---Adapted Multi-Label Silhouette Score.} Standard silhouette analysis assumes single-label clustering, but natural scene images inherently contain multiple semantic categories (e.g., ``person'' + ``bicycle'' + ``outdoor''). We implemented a modified silhouette score adapted for multi-label data:

\begin{enumerate}
\item For multi-label samples, we computed pairwise label similarity using Jaccard index:
   \begin{equation}
   J(i,j) = \frac{|\text{labels}_i \cap \text{labels}_j|}{|\text{labels}_i \cup \text{labels}_j|}
   \end{equation}
   where $\text{labels}_i$ represents the set of ground-truth categories for sample $i$.

\item For each sample $i$, we partitioned other samples into ``similar'' (Jaccard $>$ median\_similarity) and ``dissimilar'' (Jaccard $\leq$ median\_similarity) groups.

\item Distances in latent space were computed using Euclidean metric.

\item The silhouette coefficient for sample $i$ was calculated as:
   \begin{equation}
   s_i = \frac{b_i - a_i}{\max(a_i, b_i)}
   \end{equation}
   where $a_i$ = mean distance to similar samples and $b_i$ = mean distance to dissimilar samples.
\end{enumerate}
This modified silhouette score quantifies how effectively the latent space organises stimuli according to semantic category structure while accommodating the multi-label nature of natural images. For all methods, silhouette scores are computed not on the raw
$d$-dimensional latents but on a 10-dimensional UMAP embedding of the pooled multi-subject representations, fit identically across methods so that scores remain directly comparable.

\section{Methods}

\subsection{Training Objective}
\label{appendix:training_obj}
The complete MED-VAE loss function integrates the four complementary reconstruction pathways with KL regularization:

The MED-VAE loss integrates the four reconstruction pathways with KL regularization:
\begin{align}
\mathcal{L} &=
\underbrace{\mathcal{L}_{\text{fMRI}\rightarrow\text{fMRI}}}_{\text{Within-modality fMRI}}
\mkern-4mu+\mkern-4mu
\underbrace{\mathcal{L}_{\text{ANN}\rightarrow\text{fMRI}}}_{\text{Cross-modal: ANN\textrightarrow fMRI}}
\label{app:loss_eq} \\
&\quad
\mkern-4mu+\mkern-4mu
w \Big[
\underbrace{\mathcal{L}_{\text{ANN}\rightarrow\text{ANN}}}_{\text{Within-modality ANN}}
\mkern-4mu+\mkern-4mu
\underbrace{\mathcal{L}_{\text{fMRI}\rightarrow\text{ANN}}}_{\text{Cross-modal: fMRI\textrightarrow ANN}}
\Big]
\mkern-4mu+\mkern-4mu
\mathcal{L}_{\text{KL}}
\nonumber
\end{align}
where the single weighting parameter $w$ governs alignment pressure through the shared ANN decoder pathway.
and controls the alignment-reconstruction trade-off. 

\paragraph{Component 1: Within-Modality fMRI Reconstruction}

\begin{equation}
\mathcal{L}_{\text{rec}}^{\text{fMRI→fMRI}} = \sum_{i=1}^{N} \mathbb{E}_{q(\mathbf{z}_i|\mathbf{x}_i^{\text{fMRI}})} \left[\| \mathbf{x}_i^{\text{fMRI}} - \text{Dec}_i^{\text{fMRI}}(\mathbf{z}_i) \|^2 \right]
\label{eq:fmri_recon}
\end{equation}
where $q(z_i \mid x_i^{\text{fMRI}}) = \mathcal{N}(\mu_i, \text{diag}(\sigma_i^2))$ is the approximate posterior parameterised by subject $i$'s encoder (Eq.~1).
\textbf{Pathway:} $\mathbf{x}_i^{\text{fMRI}} \xrightarrow{\text{Enc}_i^{\text{fMRI}}} \mathbf{z}_i \xrightarrow{\text{Dec}_i^{\text{fMRI}}} \hat{\mathbf{x}}_i^{\text{fMRI}}$. Standard autoencoding preserves subject-specific neural information, ensuring latent representations $\mathbf{z}_i$ retain sufficient detail for within-subject voxel-level reconstruction.

\paragraph{Component 2: Cross-Modal ANN→fMRI Reconstruction}

\begin{equation}
\mathcal{L}_{\text{rec}}^{\text{ANN→fMRI}} = \sum_{i=1}^{N} \mathbb{E}_{q(\mathbf{z}_{\text{ANN}}|\mathbf{x}^{\text{ANN}})} \left[\| \mathbf{x}_i^{\text{fMRI}} - \text{Dec}_i^{\text{fMRI}}(\mathbf{z}_{\text{ANN}}) \|^2 \right]
\label{eq:ann_to_fmri}
\end{equation}
where $q(z^{\text{ANN}} \mid x^{\text{ANN}})$ by the shared ANN encoder.

\noindent \textbf{Pathway:} $\mathbf{x}^{\text{ANN}} \xrightarrow{\text{Enc}^{\text{ANN}}} \mathbf{z}_{\text{ANN}} \xrightarrow{\text{Dec}_i^{\text{fMRI}}} \hat{\mathbf{x}}_i^{\text{fMRI}}$. Enforces decoder universality by requiring subject-specific fMRI decoders to reconstruct neural activity from ANN-encoder outputs.

\paragraph{Component 3: Within-Modality ANN Reconstruction}

\begin{equation}
\mathcal{L}_{\text{rec}}^{\text{ANN→ANN}} = \mathbb{E}_{q(\mathbf{z}_{\text{ANN}}|\mathbf{x}^{\text{ANN}})} \left[\| \mathbf{x}^{\text{ANN}} - \text{Dec}^{\text{ANN}}(\mathbf{z}_{\text{ANN}}) \|^2 \right]
\label{eq:ann_autoencoding}
\end{equation}
\textbf{Pathway:} $\mathbf{x}^{\text{ANN}} \xrightarrow{\text{Enc}^{\text{ANN}}} \mathbf{z}_{\text{ANN}} \xrightarrow{\text{Dec}^{\text{ANN}}} \hat{\mathbf{x}}^{\text{ANN}}$. Establishes task-relevant latent space organization – hierarchical visual features, categorical boundaries, semantic relationships – constituting the target representational geometry for cross-subject alignment.

\paragraph{Component 4: Cross-Modal fMRI→ANN Reconstruction}
\begin{equation}
\mathcal{L}_{\text{rec}}^{\text{fMRI→ANN}} = \sum_{i=1}^{N} \mathbb{E}_{q(\mathbf{z}_i|\mathbf{x}_i^{\text{fMRI}})} \left[\| \mathbf{x}^{\text{ANN}} -\text{Dec}^{\text{ANN}}(\mathbf{z}_i) \|^2 \right]
\label{eq:fmri_to_ann}
\end{equation}
\textbf{Pathway:} $\mathbf{x}_i^{\text{fMRI}} \xrightarrow{\text{Enc}_i^{\text{fMRI}}} \mathbf{z}_i \xrightarrow{\text{Dec}^{\text{ANN}}} \hat{\mathbf{x}}^{\text{ANN}}$. Enforces encoder alignment: subject-specific fMRI encoders must produce latent codes $\mathbf{z}_i$ compatible with the shared ANN decoder. For subjects $i$ and $j$ viewing identical/similar stimuli, both $\mathbf{z}_i$ and $\mathbf{z}_j$ must reconstruct the same/similar ANN features.

\paragraph{Component 5: KL Divergence Regularization}

\begin{equation}
\mathcal{L}_{\text{KL}} = \sum_{i=1}^{N} D_{\text{KL}}(q(\mathbf{z}_i|\mathbf{x}_i^{\text{fMRI}}) \| p(\mathbf{z})) + D_{\text{KL}}(q(\mathbf{z}_{\text{ANN}}|\mathbf{x}^{\text{ANN}}) \| p(\mathbf{z}))
\label{eq:kl_divergence}
\end{equation}
where $p(\mathbf{z}) = \mathcal{N}(\mathbf{0}, \mathbf{I})$ denotes the standard Gaussian prior.

\subsection{Model Architecture}
\label{appendix:model_arch}

Each encoder is a two-layer MLP with LayerNorm, ReLU activations, and dropout ($p$). The encoder produces $\mu$ and $\log \sigma^2$ via two parallel projection heads from the final hidden layer. Each decoder mirrors this structure, with output $s \cdot \tanh(\cdot)$, where $s$ is a learnable scale parameter. Table~\ref{tab:arch} specifies the full layer sequence.

\begin{table}[h]
\centering
\small
\caption{Encoder and decoder architecture. LN = LayerNorm, D($p$) = Dropout.}
\label{tab:arch}
\begin{tabular}{cl}
\toprule
\multicolumn{2}{l}{\textbf{Encoder}} \\
\midrule
1 & Linear($d_{\text{in}}$, $2d_h$) → LN → ReLU → D($p$) \\
2 & Linear($2d_h$, $d_h$) → LN → ReLU → D($p$) \\
$\mu$ & Linear($d_h$, $d_z$) \\
$\log\sigma^2$ & Linear($d_h$, $d_z$) \\
\midrule
\multicolumn{2}{l}{\textbf{Decoder}} \\
\midrule
1 & Linear($d_z$, $d_h$) → LN → ReLU → D($p$) \\
2 & Linear($d_h$, $2d_h$) → LN → ReLU → D($p$) \\
out & Linear($2d_h$, $d_{\text{out}}$) → $s \cdot \tanh(\cdot)$ \\
\bottomrule
\end{tabular}
\end{table}

All encoders and decoders share this architecture but have independent weights.  During training, latent samples are drawn via $z = \mu + \exp(0.5 \log \sigma^2)\,\varepsilon$, $\varepsilon \sim \mathcal{N}(0, I)$; at evaluation, $z = \mu$.

\paragraph{Hyperparameters.}
We report results at $d_z = 32$, $d_h = 256$, and $p = 0.3$. For the experiments where $d_z = 512$ (Section~5.5), $d_h = 1024$. Models were trained for 30 epochs with Adam (lr $= 10^{-4}$, batch size $128$). Loss weights: $w = w_{\text{fMRI}\to\text{ANN}} = w_{\text{ANN}\to\text{ANN}} = 5$, $w_{\text{fMRI}\to\text{fMRI}} = w_{\text{ANN}\to\text{fMRI}} = 1$.

\section{Control Analysis: Untrained ANN}
\label{app:control}

To verify that cross-subject alignment in MED-VAE depends on meaningful ANN representations, we trained two control cases; one with untrained ResNet-50 and one with destroying the 1:1 correspondence between the ResNet representations and the fMRI data. In the untrained control, the ANN encoder received activations from a randomly initialised ResNet-50, which carries no learned semantic information. In the shuffled control, the ANN encoder received activations from a fully trained ResNet-50, but the mapping between stimuli and activations was randomly permuted, breaking the one-to-one correspondence between what a subject viewed and the ANN features fed to the model.

\begin{figure}[ht]
    \centering
    \includegraphics[width=0.99\columnwidth]{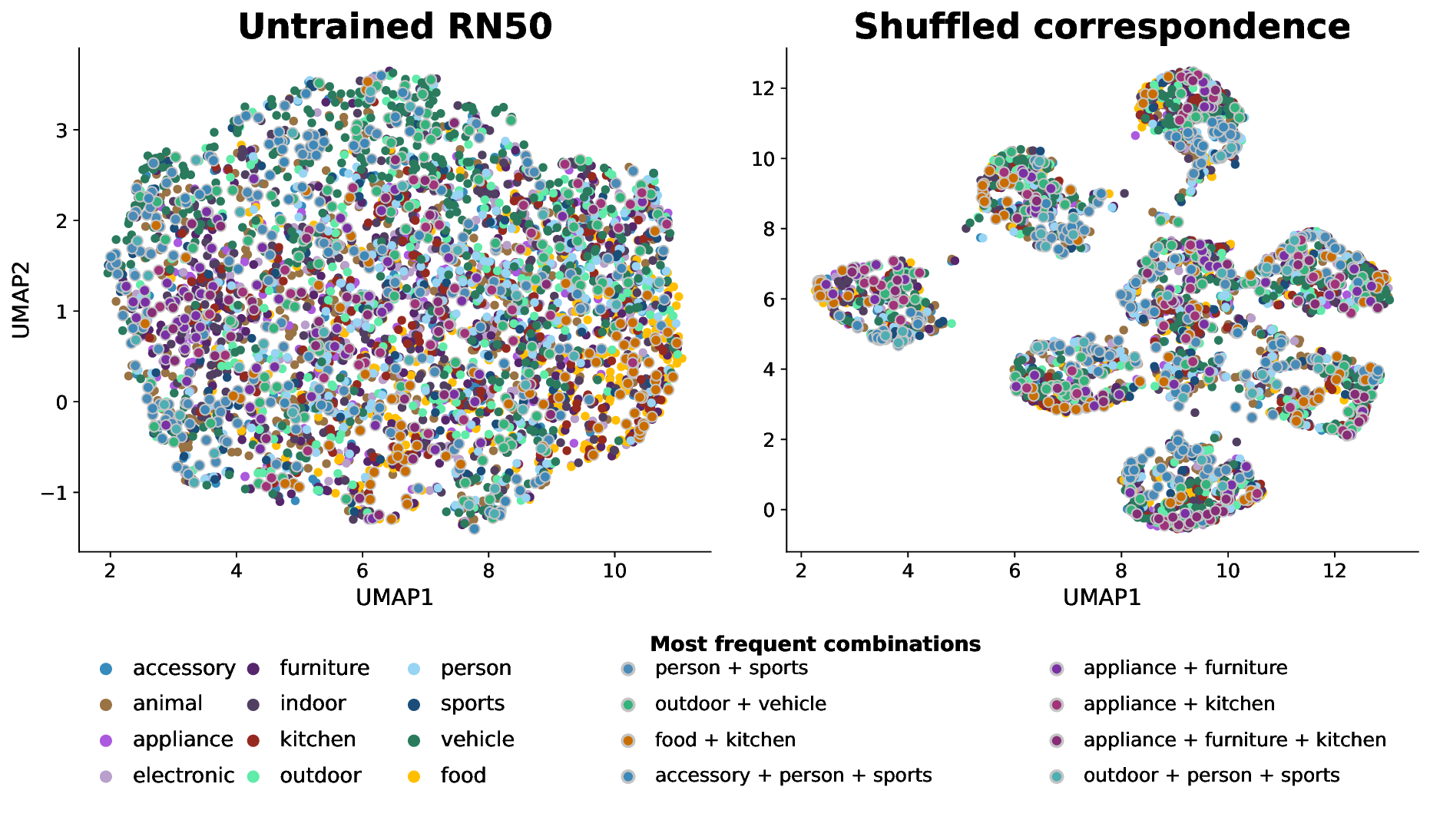}
    \caption{UMAP projection of latent representations from control MED-VAE models. (Left) Untrained MED-VAE: the ANN encoder receives activations from a randomly initialised (untrained) ResNet-50, providing no meaningful semantic signal. (Right) Shuffled MED-VAE: the ANN encoder receives activations from a trained ResNet-50, but the stimulus-to-activation mapping is randomly permuted, destroying the correspondence between ANN and fMRI responses. Points are coloured by COCO super-category label; the absence of category structure within clusters confirms that cross-subject alignment depends on a valid ANN–fMRI correspondence.}
    \label{fig:untrained}
\end{figure}

In the shuffled condition, the UMAP projection of the 32-dimensional latent space reveals eight clearly separated clusters, each corresponding to a single subject as seen in Fig. \ref{fig:untrained} (right). Because the fMRI-to-fMRI reconstruction pathway operates independently within each subject, each encoder still learns a compact representation of its own subject's data. However, without a valid shared ANN signal to anchor these representations to one another, the per-subject latent spaces remain unaligned. Consequently, no category structure emerges across subjects: points from different subjects that depict the same semantic category are scattered across separate clusters rather than co-localised as in the trained model (see. Figure \ref{fig:combined}A).
For the untrained RN50 condition, in Fig. \ref{fig:untrained} (left), we observe
one diffuse blob with thoroughly mixed colours, i.e., the latent has no semantic-category organization at all This is expected as untrained ANN features carry no category structure to align to.
Quantitatively, silhouette score in MED-VAE latent is 
0.056 for the untrained RN50 control and 0.018 for the shuffled control (trained RN50 MED-VAE is 0.34). Retrieval top-1 score is 6.6\% for the untrained control and 1.0\% for the shuffled control, compared to 46.2\% for trained RN50 MED-VAE in Table 
\ref{tab:retrieval}.

This confirms that cross-subject alignment requires the combination of the VAE architecture with a scaffold carrying meaningful computational structure; neither the VAE's architectural constraints alone nor the ResNet architecture without task-relevant learned representations is sufficient to achieve alignment.

\section{Noise Ceiling Analysis}
\label{app:noise}

\subsection{Theoretical Framework}

We model each fMRI trial as a sum of signal and noise:
\begin{equation}
Y_{\text{trial}} = S + N
\end{equation}
where $S$ represents the true stimulus-driven neural signal (consistent across trials of the same stimulus) and $N$ represents trial-specific noise. We assume $S$ and $N$ are uncorrelated: $\text{cov}(S, N) = 0$.

\paragraph{NCSNR metric.} The Natural Scenes Dataset provides per-voxel noise ceiling signal-to-noise ratio (ncsnr), defined as:
\begin{equation}
\text{ncsnr} = \frac{\sigma_S}{\sigma_N} = \frac{\sqrt{\text{var}(S)}}{\sqrt{\text{var}(N)}}
\end{equation}
This is a single scalar per voxel, estimated across all images in the dataset and reflecting the voxel's overall signal quality. Different subjects and ROIs yield different average ncsnr values. 
\paragraph{Noise ceiling as signal variance fraction.} The noise ceiling (NC) represents the proportion of total variance attributable to signal:
\begin{equation}
\text{NC} = \frac{\text{var}(S)}{\text{var}(S) + \text{var}(N)} = \frac{\text{ncsnr}^2}{\text{ncsnr}^2 + 1}
\end{equation}

\paragraph{Trial-to-trial correlation equals NC.} When correlating two independent trials of the same stimulus, both containing independent noise realisations:
\begin{align}
\text{corr}(\text{trial}_1, \text{trial}_2) &= \text{corr}(S + N_1, S + N_2) \nonumber \\
&= \frac{\text{cov}(S + N_1, S + N_2)}{\text{std}(S + N_1) \cdot \text{std}(S + N_2)} \nonumber \\
&= \frac{\text{var}(S)}{\text{var}(S) + \text{var}(N)} = \text{NC}
\end{align}
since $\text{cov}(S, N_i) = 0$ and $\text{cov}(N_1, N_2) = 0$ for independent noise.

\paragraph{Correlation ceiling.} When correlating a perfect signal prediction (noise-free) with noisy data:
\begin{align}
\text{corr}(S, S + N) &= \frac{\text{var}(S)}{\sqrt{\text{var}(S)} \cdot \sqrt{\text{var}(S) + \text{var}(N)}} \nonumber \\
&= \sqrt{\frac{\text{var}(S)}{\text{var}(S) + \text{var}(N)}} = \sqrt{\text{NC}} = R_{\text{ceil}}
\end{align}
The square root arises because only one side contains noise. This $R_{\text{ceil}}$ represents the theoretical maximum correlation achievable by any model producing perfect signal estimates. 


\paragraph{Effect of trial averaging.} When averaging $n$ trials, signal remains constant while noise variance reduces:
\begin{equation}
\text{NC}_n = \frac{\text{ncsnr}^2}{\text{ncsnr}^2 + 1/n}
\end{equation}

\subsection{Expected Gap Between Within-Trial and Cross-Trial Within-Subject Reconstruction}
\label{app:cross_trial_ceiling}

The within-trial and cross-trial reconstruction metrics reported in Fig.~\ref{fig:combined}C (Reconstruction panel) differ in their evaluation protocol, and consequently face different noise ceilings. Here, we derive the expected gap between them for the ncsnr value of the ROIs we used in our analysis, which is provided by NSD and shown in Table~\ref{tab:nc-within-cross-trial}, to verify that the observed difference in values is consistent with the noise ceiling framework rather than reflecting a loss of neural information.

\paragraph{Within-trial evaluation.} In our within-trial protocol, we reconstruct from trial-averaged input ($n = 3$ repetitions) and correlate the reconstruction $\hat{S}$ with the same trial-averaged ground truth $S + N_{\text{avg}}$, where $\text{var}(N_{\text{avg}}) = \text{var}(N)/3$. For a model that perfectly recovers stimulus-driven signal, the ceiling is:
\begin{equation}
R_{\text{ceil},n=3} = \sqrt{\text{NC}_{n=3}} = \sqrt{\frac{\text{ncsnr}^2}{\text{ncsnr}^2 + 1/3}}
\end{equation}
The values for each subject's $R_{\text{ceil},n=3}$ are given in Table~\ref{tab:nc-within-cross-trial}, with mean $R_{\text{ceil},n=3}=0.618$.

\paragraph{Cross-trial evaluation.} In our cross-trial protocol, we reconstruct from a single trial and correlate the reconstruction with an independent single trial of the same stimulus. Even if the model perfectly extracts signal from its input trial, the evaluation target $S + N_j$ contains full single-trial noise ($\text{var}(N_j) = \text{var}(N)$), so the ceiling is:
\begin{equation}
R_{\text{ceil},n=1} = \sqrt{\text{NC}_{n=1}} = \sqrt{\frac{\text{ncsnr}^2}{\text{ncsnr}^2 + 1}}
\end{equation}
As shown in Table~\ref{tab:nc-within-cross-trial},  mean $R_{\text{ceil},n=1} = 0.449$.
\setlength{\tabcolsep}{2pt}
\begin{table}[h]
\centering
\small
\caption{Per-subject mean NCSNR, noise ceilings, and correlation ceilings, averaged across all $\sim$20{,}730 occipitotemporal cortex (Streams ROI) voxels. $K{=}3$: within-trial (trial-averaged) evaluation. $K{=}1$: cross-trial (single-trial) evaluation.}
\label{tab:nc-within-cross-trial}
\begin{tabular}{lccccc}
\toprule
\textbf{Subject} & \textbf{NCSNR} & \textbf{NC ($K{=}3$)} & \textbf{$R_{\text{ceil}}$ ($K{=}3$)} & \textbf{NC ($K{=}1$)} & \textbf{$R_{\text{ceil}}$ ($K{=}1$)} \\
\midrule
S1 & 0.501 & 0.407 & 0.622 & 0.201 & 0.448 \\
S2 & 0.538 & 0.437 & 0.640 & 0.224 & 0.474 \\
S5 & 0.576 & 0.464 & 0.662 & 0.249 & 0.499 \\
S7 & 0.406 & 0.323 & 0.549 & 0.142 & 0.376 \\
\midrule
\textbf{Overall} & \textbf{0.505} & \textbf{0.408} & \textbf{0.618} & \textbf{0.204} & \textbf{0.449} \\
\bottomrule
\end{tabular}
\end{table}

\paragraph{Summary.} We denote by $\hat{S}$ the model's reconstruction from a given input.
\begin{center}
\fontsize{8}{9}\selectfont
\begin{tabular}{llc}
\toprule
\textbf{Evaluation} & \textbf{Correlation computed} & \textbf{Ceiling} \\
\midrule
Within-trial & $\text{corr}(\hat{S}_{\text{avg}},\; S + N_{\text{avg}})$ & $R_{\text{ceil},n=3} = 0.618$ \\
Cross-trial & $\text{corr}(\hat{S}_{\text{trial}_i},\; S + N_j), \;\; j \neq i$ & $R_{\text{ceil},n=1} = 0.449$ \\
Baseline & $\text{corr}(S + N_i,\; S + N_j), \;\; j \neq i$ & $\text{NC}_{n=1} = 0.204$ \\
\bottomrule
\end{tabular}
\end{center}

The within-trial ceiling ($R_{\text{ceil},n=3} = 0.618$) is substantially higher than the cross-trial ceiling ($R_{\text{ceil},n=1} = 0.449$) for two reasons: (i) the within-trial evaluation target contains noise reduced by $n=3$ averaging, and (ii) the model reconstructs from this less noisy averaged input to itself. The cross-trial ceiling of $0.447$ would be reached only if the model produced a perfectly denoised signal estimate from a single trial.

In practice, all methods achieve cross-trial correlations of approximately $0.20$ (Fig.~\ref{fig:combined}C, Reconstruction panel), coinciding with the no-model baseline $\text{NC}_{n=1} \approx 0.20$. The key observation is that all methods---MED-VAE, SRM, and Procrustes---converge to the same cross-trial performance despite within-trial values ranging from $\approx 0.6$ (MED-VAE) to $\approx 0.7$ (SRM). The entire within-trial gap between methods vanishes under cross-trial evaluation, confirming that it reflects non stimulus-driven signal.

\subsection{NC-Normalised Within-Subject Reconstruction}
\label{app:nc_norm}

The cross-trial analysis above demonstrates that all methods preserve equal stimulus-driven signal. To further quantify the extent to which each method's within-trial reconstruction reflects signal versus noise, we normalise self-reconstruction performance by the per-voxel noise ceiling, that the dataset provides.
For each voxel $v$, we compute:
\begin{equation}
\text{NC-norm}_v = \frac{r_v}{R_{\text{ceil},v}}
\end{equation}
where $r_v$ is the method's Pearson correlation between reconstructed and original responses at voxel $v$ (computed across images), and $R_{\text{ceil},v} = \sqrt{\text{NC}_v}$ is the per-voxel correlation ceiling computed from that voxel's ncsnr with $K{=}3$ trial averaging. We then average across all voxels:
\begin{equation}
\text{NC-norm} = \frac{1}{V} \sum_{v=1}^{V} \frac{r_v}{R_{\text{ceil},v}}
\end{equation}
expressed as a percentage. This per-voxel normalisation accounts for heterogeneous signal quality across voxels. Values above 100\% indicate that the method reconstructs more variance than can be attributed to stimulus-driven signal at individual voxels, i.e., it captures trial-specific noise shared between input and evaluation target. 

Table~\ref{tab:nc-norm-methods} reports NC-normalised performance for all three alignment methods on the 128 held-out images.

\setlength{\tabcolsep}{0.5pt} 
\begin{table}[h]

\centering
\small
\caption{Within-subject reconstruction and cross-subject prediction with NC-normalisation. NC-norm values above 100\% on the diagonal (within-subject reconstruction) indicate reconstruction of non-stimulus-driven variance.}
\label{tab:nc-norm-methods}
\begin{tabular}{lccc}
\toprule
\textbf{Metric} & \textbf{Procrustes} & \textbf{SRM} & \textbf{MED-VAE} \\
\midrule
within-subj recon $r$ (within-trial) & 0.746 & 0.707 & 0.606 \\
NC-norm (diagonal)            & 154.8\% & 138.3\% & 102.5\% \\
\midrule
Cross-subj prediction $r$     & 0.332 & 0.344 & 0.383 \\
NC-norm (off-diagonal)        & 33.4\% & 35.4\% & 40.9\% \\
\bottomrule
\end{tabular}
\end{table}

\paragraph{Interpretation.} The NC-normalised within-subject reconstruction values confirm the noise-capture interpretation from the cross-trial analysis:

Procrustes (154.8\%) and SRM (138.3\%) substantially exceed the noise ceiling. This confirms that their within-trial within-subject reconstruction advantage derives from capturing trial-specific measurement noise, i.e., spatially correlated noise patterns that inflate within-trial correlations but vanish under cross-trial evaluation.

Overall, cross-trial analysis and NC-normalised values demonstrate that MED-VAE does not sacrifice stimulus-driven signal for alignment. Its lower within-trial reconstruction relative to classical methods reflects principled noise suppression, not inferior signal capture.

\section{Hyperparameter $w$ Sensitivity}
\label{app:w_sweep} 

We swept $w \in \{0.5, 1, 2, 5, 10, 20\}$ on the 872 shared images (held out from training). Table~\ref{tab:w_sweep} reports key metrics.

\begin{table}[h]
\centering
\small
\caption{Sensitivity to scaffold weight $w$. Selected value in bold.}
\label{tab:w_sweep}
\begin{tabular}{ccccccc}
\toprule
$w$ & CompCorr & Ret@1 & Silh. & $r_{\text{vox}}$ (cross) & $r_{\text{vox}}$ (self) & NC-norm \\
\midrule
0.5 & 0.558 & 14.5\% & 0.331 & 0.374 & 0.657 & 131.6\% \\
1   & 0.584 & 17.0\% & 0.336 & 0.376 & 0.647 & 127.8\% \\
2   & 0.615 & 18.9\% & 0.352 & 0.379 & 0.633 & 122.0\% \\
\textbf{5} & \textbf{0.655} & \textbf{19.6\%} & \textbf{0.358} & \textbf{0.383} & 0.610 & 112.9\% \\
10  & 0.671 & 18.8\% & 0.348 & 0.384 & 0.590 & 105.6\% \\
20  & 0.682 & 17.9\% & 0.340 & 0.382 & 0.568 & 97.3\% \\
\bottomrule
\end{tabular}
\end{table}

Increasing $w$ monotonically improves  component-wise correlation (CompCorr) but degrades within-subject reconstruction correlation
($r_{\text{vox}}$ (self)). Noise-ceiling-normalised within-reconstruction
(NC-norm; Appendix~\ref{app:nc_norm}) remains above 100\% for $w \leq 10$, dropping below only at $w = 20$. Downstream alignment metrics cross-subject retrieval top-1 accuracy  (Ret@1) and silhouette score peak at $w = 5$ and decline at higher values, while cross-subject voxel prediction correlation
($r_{\text{vox}}$) is stable across the sweep (0.374--0.384). We selected $w = 5$ as the highest value maintaining NC-norm above 100\% while maximising downstream alignment.

\section{Per-Layer Reconstruction Analysis}
\label{app:per-layer}

To understand what the shared latent space captures across the scaffold hierarchy, we performed a per-layer analysis evaluating two reconstruction pathways: ANN$\to$latent$\to$ANN (how much of each layer's information survives the 32D bottleneck) and fMRI$\to$latent$\to$ANN (how much of each layer's information is mapped to brain data, and thus reconstructed by the fMRI$\to$latent$\to$ANN pathway). All values are per-feature Pearson correlations.

Raw fMRI$\to$ANN reconstruction is confounded by each ANN layer's intrinsic dimensionality: layers with high intrinsic dimensionality are disfavoured by the low-dimensional bottleneck regardless of their brain relevance, simply because their variance cannot be compressed into 32 dimensions. We therefore report the \textbf{ratio} of fMRI$\to$ANN to ANN$\to$ANN correlation as the primary metric, quantifying how each layer maps to the fMRI data. This ratio controls for compressibility and indexes brain--ResNet-50 alignment at each hierarchical level of ResNet-50: it measures what fraction of each layer's preserved information (i.e., the information that survives the bottleneck) is neurally accessible, i.e., maps to the fMRI data.

For the ResNet-50 data, we sampled one layer every 8 (both pre- and post-activation), yielding 15 sampled layers spanning the full depth of the network. Each layer's activations were reduced to 3{,}500 dimensions via Sparse Random Projection before concatenation into a single 51K-dimensional feature vector. The results for all layer activations, are the following: 

\setlength{\tabcolsep}{1pt} 

\setlength{\tabcolsep}{1pt} 
\begin{table}[h]
\centering
\small
\caption{Per-layer reconstruction for ResNet-50 (MED-VAE, $w=5$). We have included activations from Conv2d layers (pre-ReLU) and post-ReLU. Per-feature Pearson correlation on 872 shared images. \textbf{Ratio} = fMRI$\to$ANN / ANN$\to$ANN.}
\label{tab:perlayer-rn50}
\begin{tabular}{llccc}
\toprule
\textbf{Layer} & \textbf{Module} & \textbf{ANN$\to$ANN} & \textbf{fMRI$\to$ANN} & \textbf{Ratio} \\
\midrule

Conv2d-1  & conv1   & 0.514 & 0.263 & 0.51 \\
ReLU-1    & conv1   & 0.420 & 0.218 & 0.52 \\

Conv2d-9  & layer1  & 0.209 & 0.145 & 0.69 \\
ReLU-9    & layer1  & 0.141 & 0.100 & 0.71 \\

Conv2d-17 & layer2  & 0.455 & 0.259 & 0.57 \\
ReLU-17   & layer2  & 0.322 & 0.231 & 0.72 \\

Conv2d-25 & layer3  & 0.225 & 0.161 & 0.71 \\
ReLU-25   & layer3  & 0.339 & 0.240 & 0.71 \\

Conv2d-33 & layer3  & 0.335 & 0.253 & 0.76 \\
ReLU-33   & layer3  & 0.344 & 0.256 & 0.74 \\

Conv2d-41 & layer3  & 0.409 & 0.315 & 0.77 \\
ReLU-41   & layer3  & 0.292 & 0.204 & 0.70 \\

Conv2d-49 & layer4  & 0.534 & 0.428 & 0.80 \\
ReLU-49   & layer4  & 0.464 & 0.363 & 0.78 \\

\midrule
AvgPool   & final avgpool & 0.586 & 0.469 & 0.80 \\
\bottomrule
\end{tabular}
\end{table}

Table~\ref{tab:perlayer-rn50} reveals three distinct patterns across the ResNet-50 hierarchy:

\paragraph{ANN$\to$ANN reconstruction is $\cup$-shaped.} Early layers (conv1: 0.514) and late layers (avgpool: 0.586) reconstruct best, while intermediate layers show a pronounced trough. This profile mirrors the inverse of the known intrinsic dimensionality (ID) profile across CNN layers: \citet{ansuini2019intrinsic} showed that ID peaks at intermediate ResNet layers before dropping again for the final layers.


\paragraph{The brain--RN50 ratio increases monotonically with depth.} The ratio rises from 0.51 at conv1 to 0.80 at layer4/avgpool. At every successive level of the hierarchy, a larger fraction of what the bottleneck preserved is brain-accessible, i.e., maps to the fMRI data. 
This is consistent with the functional properties of our occipitotemporal ROIs (LOC, FFA, PPA, EBA), which encode high-level categorical and object-selective representations corresponding to late network layers.

For a compact summary, Table~\ref{tab:depth-rn50} aggregates across pre- and post-ReLu activation layers at three depth levels.

\begin{table}[h]
\centering
\small
\caption{Reconstruction summary at different ResNet-50 depths summary (averaging pre- and post-activation layers).}
\label{tab:depth-rn50}
\begin{tabular}{lccc}
\toprule
\textbf{Depth} & \textbf{ANN$\to$ANN} & \textbf{fMRI$\to$ANN} & \textbf{Ratio} \\
\midrule
Early (conv1, layer1) & 0.321 & 0.182 & 0.57 \\
Mid (layer2, layer3)  & 0.340 & 0.240 & 0.71 \\
Late (layer4, avgpool) & 0.528 & 0.420 & 0.80 \\
\bottomrule
\end{tabular}
\end{table}

\section{Scaffold Comparison: ResNet-50 vs CLIP ViT-L/14}
\label{app:scaffold}

To assess the sensitivity of MED-VAE's alignment on the use of different scaffold architectures, we trained the full framework with two distinct scaffolds: ResNet-50 (ImageNet-pretrained, $\sim$51K features after per-layer SRP) and CLIP ViT-L/14 (contrastive vision transformer, $\sim$26K features after per-layer SRP). In both cases, MED-VAE uses a 32-dimensional latent bottleneck and, here, evaluations are on the 872 shared images across all 8 NSD subjects (held-out from training).

Table~\ref{tab:scaffold-alignment} reports the full comparison across all alignment, reconstruction, and downstream metrics.
\begin{table}[h]
\centering
\small
\caption{Full scaffold comparison across all evaluation metrics. Bold indicates the better-performing scaffold. Evaluated on 872 shared images, all 8 NSD subjects.}
\label{tab:scaffold-alignment}
\begin{tabular}{lcc}
\toprule
\textbf{Metric} & \textbf{ResNet-50} & \textbf{CLIP ViT-L/14} \\
\midrule
\multicolumn{3}{l}{\textit{Latent space alignment}} \\
Compositional Corr. & \textbf{0.647} & 0.549 \\
RSA (Pearson) & \textbf{0.652} & 0.582 \\
\midrule
\multicolumn{3}{l}{\textit{Category encoding}} \\
Silhouette & 0.336 & \textbf{0.353} \\
Balanced Accuracy & \textbf{0.832} & 0.809 \\
\midrule
\multicolumn{3}{l}{\textit{Cross-subject retrieval}} \\
Retrieval Top-1 & \textbf{48.9\%} & 31.4\% \\
Retrieval Top-2 & \textbf{63.6\%} & 48.1\% \\
\midrule
\multicolumn{3}{l}{\textit{Reconstruction}} \\
fMRI Recon (voxel corr.) & 0.606 & 0.606 \\
fMRI Recon (NC norm\%) & 102.5 & 102.7 \\
Cross-Trial $r$ & \textbf{0.269} & 0.257 \\
\midrule
\multicolumn{3}{l}{\textit{Cross-subject neural prediction}} \\
Cross-Subj Prediction & \textbf{0.383} & 0.370 \\
Cross-Subj NC\% & \textbf{40.8} & 38.6 \\
\bottomrule
\end{tabular}
\end{table}
ResNet-50 outperforms CLIP ViT-L/14 on the majority of cross-subject alignment metrics, with particularly large advantages in retrieval accuracy (48.9\% vs 31.4\% top-1) and latent alignment metrics (CompCorr: 0.647 vs 0.549).
Importantly though, MED-VAE with CLIP ViT-L/14 still substantially outperforms all classical baselines benchmarked in the paper (SRM, Procrustes), showcasing a robustness to the choice of the ANN scaffold.

\section{ANN Reconstruction Performance}
\label{app:ann_recon}

Here we explore the ANN reconstruction performance of the ANN$\rightarrow$ANN pathway measured by Pearson correlation between the original activations and the reconstructed ones through the pathway.
To isolate the cost of sharing the latent space with fMRI, we compare MED-VAE, where the latent is trained on ANN and fMRI concurrently,  against an ANN-only VAE, i.e., same 32D architecture, trained with only the ANN encoder/decoder --- no fMRI encoders or decoders. Table~\ref{tab:ann-recon} reports per-feature Pearson correlation on the 128 held-out images.

\begin{table}[h]
\centering
\small
\caption{ANN reconstruction: MED-VAE vs ANN-only control.}
\label{tab:ann-recon}
\begin{tabular}{lc}
\toprule
\textbf{Model} & \textbf{Per-feature $r$} \\
\midrule
ANN-only VAE & 0.368 \\
MED-VAE ($w{=}5$) & 0.354 \\
\bottomrule
\end{tabular}
\end{table}

The $\sim$4\% difference confirms that sharing the latent space with fMRI does not reduce the ANN reconstruction performance; the 32D bottleneck compressing 51K ResNet-50 features is the binding constraint, not the trade-off with fMRI pathways.
A PCA analysis of the ResNet-50 features explains this constraint: 32 PCs capture only 14.6\% of the total variance, and $\sim$1{,}810 PCs are needed to reach 50\% --- reflecting the high intrinsic dimensionality of intermediate ResNet layers \citep{ansuini2019intrinsic}. Per layer ANN$\rightarrow$ANN reconstruction, along with fMRI$\rightarrow$ANN reconstruction, i.e., how much of each layer’s reconstructed information is mapped to the fMRI data, can be found in Appendix~\ref{app:per-layer}.



 \section{Generalisation to Novel Subjects}
\label{app:novel-subjects}
A practical requirement for cross-subject alignment is that a new participant can
be added to an existing shared space efficiently---ideally without retraining the
space and without requiring stimuli shared with the existing cohort. We assess
this in two ways: (i) a leave-one-subject-out integration experiment testing
whether a held-out subject can be aligned to a frozen shared space, and (ii) a
data-efficiency analysis quantifying how much of the new participant's data is
needed to achieve good alignment.
\subsection{Leave-one-subject-out integration}
\label{app:loso-integration}
We trained MED-VAE on 7 of the 8 NSD subjects to shape the shared latent space,
then integrated the held-out subject by training only a new subject-specific
encoder and decoder to align with the ANN scaffold, keeping the shared ANN
encoder and decoder---and hence the shared space itself---frozen. This was
repeated with each of the 8 subjects held out in turn.
Crucially, the new subject's encoder and decoder are trained on that subject's own
(non-overlapping) stimuli; no stimulus overlap with the other 7 subjects is
required, and only an encoder--decoder pair is fitted, making integration
computationally cheap.
We compared this finetuned integration against the full jointly-trained 8-subject
model (MED-VAE trained with all 8 subjects jointly), evaluating both on the 872
images shared across all subjects (held out from MED-VAE training).
Table~\ref{tab:loso} reports performance averaged over all held-out subjects.
\begin{table}[H]
\centering
\small
\setlength{\tabcolsep}{4pt}
\caption{Leave-one-subject-out integration into a frozen shared space vs.\ the
full jointly-trained 8-subject model. Values are averaged across all held-out
subjects, evaluated on the 872 shared images.}
\label{tab:loso}
\begin{tabular}{lcc}
\toprule
Metric & Joint & Finetuned \\
\midrule
Compositional Corr.      & 0.664   & 0.655   \\
RSA (Pearson)            & 0.656   & 0.638   \\
Retrieval Top-1 (\%)     & 18.1    & 17.8    \\
Self-recon.\ (voxel $r$) & 0.624   & 0.587   \\
Self-recon.\ (NC\%)      & 118.9\% & 104.1\% \\
\bottomrule
\end{tabular}
\end{table}
Finetuning into the frozen shared space recovers approximately 98\% of the
jointly-trained component correlation, RSA, and retrieval accuracy. The raw voxel
reconstruction correlation drops by 0.037 on average, but noise-ceiling
normalisation (Appendix~\ref{app:nc_norm}) shows this does not reflect a loss of
stimulus-driven signal: NC-normalised reconstruction remains above 100\%, i.e.\ the finetuned model continues to recover essentially all of the
recoverable stimulus-driven variance. MED-VAE therefore provides a simple route
to integrate novel subjects, achieving performance comparable to including the
subject in the initial joint training.
\emph{(Metrics here are measured on the 872 shared images and so
differ from the 128-image values reported in the main text.)}
\subsection{Data efficiency}
\label{app:data-efficiency}
To quantify how much data a new participant needs in order to align to the group
that defines the shared space, we used subject 5 as a representative held-out subject: we
trained MED-VAE on the remaining 7 subjects, froze the shared ANN encoder and
decoder, and then trained a new encoder and decoder for subject 5 on 10\%, 20\%, 30\%,
and 100\% of its available data. Table~\ref{tab:data-efficiency} reports the
results.
\begin{table}[H]
\centering
\footnotesize
\setlength{\tabcolsep}{4pt}
\caption{Data efficiency of integrating a held-out subject (S5) into a frozen
shared space, as a function of the fraction of S5 data used to train its
encoder--decoder pair. Self-recon.\ columns report within-subject reconstruction.}
\label{tab:data-efficiency}
\begin{tabular}{lcccc}
\toprule
     & Comp. & RSA      & \multicolumn{2}{c}{Self-recon.} \\
\cmidrule(lr){4-5}
Data & Corr. & (Pears.) & voxel $r$ & NC\% \\
\midrule
10\%  & 0.655 & 0.664 & 0.526 & 63.9 \\
20\%  & 0.661 & 0.662 & 0.621 & 89.6 \\
30\%  & 0.668 & 0.672 & 0.632 & 92.9 \\
100\% & 0.679 & 0.672 & 0.655 & 99.5 \\
\bottomrule
\end{tabular}
\end{table}

The latent alignment metrics are remarkably robust to data reduction: at just 10\%
of the training data, component correlation reaches 96.5\% of the full-data value
and RSA reaches 98.8\%. Self-reconstruction shows the largest improvement between
10\% and 20\% of the data (0.526$\rightarrow$ 0.621, i.e. 64\% $\rightarrow$
90\% of the noise ceiling), with diminishing returns thereafter. Together these
results indicate that a new participant can be aligned to an existing,
well-scanned cohort with substantially less data, without shared stimuli and
without retraining the shared representational space---a regime directly relevant
to clinical settings where scanning time is constrained.

\onecolumn

\section{Cross-Subject Image Decoding Results: Per-Pair Results}
\label{Appendix:image_decoding}

Tables~\ref{tab:tab_vae_srm_proc_32d_norefine} and \ref{tab:tab_vae_srm_proc_512d_norefine} present cross-subject image decoding performance without refinement, while Tables~\ref{tab:vae_srm_32d_refine} and \ref{tab:vae_srm_512d_refine} present results with refinement, for 32-dimensional and 512-dimensional common spaces, respectively. Refinement refers to the second stage of the MindEye2 pipeline \citep{scotti2023reconstructing}, in which initial reconstructions generated via SDXL unCLIP are passed through a base Stable Diffusion XL image-to-image process conditioned on predicted text captions, improving perceptual quality without substantially altering high-level semantic content. Notation  $i \rightarrow j$ denotes alignment of subject
$i$'s neural responses to subject
$j$'s space, with subsequent decoding via a frozen MindEye2 decoder. For each subject pair and metric, the superior method is indicated in bold.

\FloatBarrier

\begin{table*}[ht]
\caption{MED-VAE 32D vs SRM 32D vs Procrustes 32D decoding performance (No Refinement). Bold indicates the best-performing method per pair. Higher is better ($\uparrow$) for all metrics except Eff and SwAV ($\downarrow$).}
\label{tab:tab_vae_srm_proc_32d_norefine}
\centering
\scriptsize
\begin{tabular}{ll cccccccc cc}
\toprule
Pair & Method
& PixCorr$\uparrow$
& SSIM$\uparrow$
& Alex2$\uparrow$
& Alex5$\uparrow$
& Incep$\uparrow$
& CLIP$\uparrow$
& Eff$\downarrow$
& SwAV$\downarrow$
& Fwd\%$\uparrow$
& Bwd\%$\uparrow$ \\
\midrule

\multirow{3}{*}{$2\rightarrow1$}
& MED-VAE & \textbf{0.182} & 0.349 & \textbf{0.868} & \textbf{0.941} & \textbf{0.880} & \textbf{0.826} & \textbf{0.790} & \textbf{0.465} & \textbf{85.0} & 65.6 \\
& SRM & 0.134 & 0.350 & 0.828 & 0.906 & 0.783 & 0.746 & 0.859 & 0.517 & 78.5 & \textbf{71.0} \\
& Proc & 0.123 & \textbf{0.352} & 0.831 & 0.909 & 0.795 & 0.728 & 0.866 & 0.523 & 69.9 & 62.4 \\
\midrule

\multirow{3}{*}{$5\rightarrow1$}
& MED-VAE & \textbf{0.154} & 0.347 & \textbf{0.837} & \textbf{0.924} & \textbf{0.857} & \textbf{0.814} & \textbf{0.769} & \textbf{0.461} & \textbf{78.6} & \textbf{53.8} \\
& SRM & 0.134 & 0.345 & 0.770 & 0.874 & 0.817 & 0.737 & 0.845 & 0.511 & 64.5 & 52.7 \\
& Proc & 0.101 & \textbf{0.359} & 0.745 & 0.851 & 0.794 & 0.704 & 0.868 & 0.529 & 41.9 & 40.9 \\
\midrule

\multirow{3}{*}{$7\rightarrow1$}
& MED-VAE & \textbf{0.132} & 0.315 & \textbf{0.796} & \textbf{0.888} & \textbf{0.827} & \textbf{0.795} & \textbf{0.819} & \textbf{0.492} & \textbf{59.1} & \textbf{53.8} \\
& SRM & 0.077 & 0.321 & 0.717 & 0.805 & 0.717 & 0.693 & 0.895 & 0.554 & 45.2 & 36.2 \\
& Proc & 0.077 & \textbf{0.332} & 0.678 & 0.753 & 0.665 & 0.641 & 0.919 & 0.564 & 35.5 & 25.8 \\
\midrule

\multirow{3}{*}{$1\rightarrow2$}
& MED-VAE & \textbf{0.226} & \textbf{0.354} & \textbf{0.885} & \textbf{0.960} & \textbf{0.923} & \textbf{0.878} & \textbf{0.717} & \textbf{0.421} & \textbf{90.3} & \textbf{77.4} \\
& SRM & 0.145 & 0.347 & 0.804 & 0.893 & 0.804 & 0.774 & 0.848 & 0.509 & 78.5 & 69.9 \\
& Proc & 0.141 & 0.345 & 0.799 & 0.883 & 0.793 & 0.737 & 0.873 & 0.527 & 67.7 & 65.6 \\
\midrule

\multirow{3}{*}{$5\rightarrow2$}
& MED-VAE & \textbf{0.145} & 0.334 & \textbf{0.842} & \textbf{0.932} & \textbf{0.897} & \textbf{0.821} & \textbf{0.790} & \textbf{0.476} & \textbf{71.0} & \textbf{62.4} \\
& SRM & 0.131 & 0.344 & 0.752 & 0.859 & 0.815 & 0.746 & 0.847 & 0.527 & 61.3 & 56.3 \\
& Proc & 0.130 & \textbf{0.348} & 0.737 & 0.842 & 0.772 & 0.699 & 0.861 & 0.539 & 44.1 & 43.0 \\
\midrule

\multirow{3}{*}{$7\rightarrow2$}
& MED-VAE & \textbf{0.126} & \textbf{0.315} & \textbf{0.793} & \textbf{0.893} & \textbf{0.826} & \textbf{0.772} & \textbf{0.805} & \textbf{0.488} & \textbf{61.3} & \textbf{57.6} \\
& SRM & 0.062 & 0.312 & 0.706 & 0.805 & 0.716 & 0.699 & 0.905 & 0.564 & 41.9 & 33.3 \\
& Proc & 0.081 & 0.314 & 0.686 & 0.752 & 0.655 & 0.661 & 0.908 & 0.562 & 31.2 & 26.9 \\
\midrule

\multirow{3}{*}{$1\rightarrow5$}
& MED-VAE & \textbf{0.170} & \textbf{0.353} & \textbf{0.882} & \textbf{0.944} & \textbf{0.911} & \textbf{0.881} & \textbf{0.712} & \textbf{0.418} & \textbf{90.3} & \textbf{74.2} \\
& SRM & 0.101 & 0.342 & 0.799 & 0.864 & 0.775 & 0.716 & 0.877 & 0.528 & 76.9 & 71.0 \\
& Proc & 0.085 & 0.337 & 0.783 & 0.871 & 0.774 & 0.701 & 0.892 & 0.539 & 51.6 & 48.4 \\
\midrule

\multirow{3}{*}{$2\rightarrow5$}
& MED-VAE & \textbf{0.173} & 0.344 & \textbf{0.860} & \textbf{0.933} & \textbf{0.874} & \textbf{0.839} & \textbf{0.787} & \textbf{0.470} & \textbf{75.3} & \textbf{69.9} \\
& SRM & 0.117 & \textbf{0.355} & 0.796 & 0.894 & 0.776 & 0.732 & 0.873 & 0.527 & 69.9 & 55.9 \\
& Proc & 0.131 & 0.353 & 0.785 & 0.866 & 0.715 & 0.722 & 0.895 & 0.542 & 57.0 & 50.5 \\
\midrule

\multirow{3}{*}{$7\rightarrow5$}
& MED-VAE & \textbf{0.121} & 0.306 & \textbf{0.799} & \textbf{0.886} & \textbf{0.842} & \textbf{0.784} & \textbf{0.814} & \textbf{0.492} & \textbf{64.5} & \textbf{55.9} \\
& SRM & 0.097 & 0.311 & 0.718 & 0.809 & 0.726 & 0.682 & 0.913 & 0.558 & 36.6 & 35.5 \\
& Proc & 0.085 & \textbf{0.319} & 0.708 & 0.765 & 0.628 & 0.643 & 0.925 & 0.568 & 31.2 & 21.5 \\
\midrule

\multirow{3}{*}{$1\rightarrow7$}
& MED-VAE & \textbf{0.189} & 0.361 & \textbf{0.914} & \textbf{0.955} & \textbf{0.895} & \textbf{0.887} & \textbf{0.736} & \textbf{0.423} & \textbf{89.3} & \textbf{74.2} \\
& SRM & 0.114 & \textbf{0.366} & 0.813 & 0.870 & 0.801 & 0.768 & 0.868 & 0.521 & 79.0 & 69.9 \\
& Proc & 0.096 & 0.365 & 0.777 & 0.866 & 0.740 & 0.718 & 0.882 & 0.533 & 57.0 & 54.8 \\
\midrule

\multirow{3}{*}{$2\rightarrow7$}
& MED-VAE & \textbf{0.163} & 0.342 & \textbf{0.851} & \textbf{0.923} & \textbf{0.864} & \textbf{0.830} & \textbf{0.788} & \textbf{0.453} & \textbf{81.7} & \textbf{65.6} \\
& SRM & 0.136 & \textbf{0.367} & 0.815 & 0.890 & 0.736 & 0.742 & 0.844 & 0.503 & 68.8 & 62.4 \\
& Proc & 0.131 & \textbf{0.368} & 0.795 & 0.838 & 0.760 & 0.727 & 0.864 & 0.528 & 54.7 & 57.0 \\
\midrule

\multirow{3}{*}{$5\rightarrow7$}
& MED-VAE & \textbf{0.139} & 0.341 & \textbf{0.827} & \textbf{0.911} & \textbf{0.839} & \textbf{0.797} & \textbf{0.791} & \textbf{0.475} & \textbf{75.3} & \textbf{62.4} \\
& SRM & 0.107 & 0.345 & 0.770 & 0.873 & 0.762 & 0.738 & 0.846 & 0.523 & 69.9 & 55.9 \\
& Proc & 0.104 & \textbf{0.348} & 0.766 & 0.837 & 0.744 & 0.710 & 0.869 & 0.539 & 52.7 & 47.3 \\
\midrule\midrule

\multirow{3}{*}{\textbf{Mean}}
& MED-VAE & \textbf{0.160} & 0.338 & \textbf{0.846} & \textbf{0.925} & \textbf{0.870} & \textbf{0.827} & \textbf{0.776} & \textbf{0.463} & \textbf{76.7} & \textbf{64.4} \\
& SRM & 0.113 & 0.342 & 0.774 & 0.862 & 0.769 & 0.731 & 0.868 & 0.529 & 64.3 & 55.9 \\
& Proc & 0.107 & \textbf{0.345} & 0.758 & 0.836 & 0.736 & 0.699 & 0.885 & 0.541 & 49.5 & 45.3 \\
\bottomrule
\end{tabular}
\end{table*}

\begin{table*}[h]
\caption{MED-VAE 512D vs SRM 512D vs Procrustes 512D decoding performance (No Refinement). Bold indicates the best-performing method per pair. Higher is better ($\uparrow$) for all metrics except Eff and SwAV ($\downarrow$).}
\label{tab:tab_vae_srm_proc_512d_norefine}
\centering
\scriptsize
\begin{tabular}{ll cccccccc cc}
\toprule
Pair & Method
& PixCorr$\uparrow$
& SSIM$\uparrow$
& Alex2$\uparrow$
& Alex5$\uparrow$
& Incep$\uparrow$
& CLIP$\uparrow$
& Eff$\downarrow$
& SwAV$\downarrow$
& Fwd\%$\uparrow$
& Bwd\%$\uparrow$ \\
\midrule

\multirow{3}{*}{$2\rightarrow1$}
& MED-VAE & 0.192 & \textbf{0.366} & \textbf{0.906} & \textbf{0.966} & \textbf{0.923} & \textbf{0.854} & \textbf{0.746} & \textbf{0.431} & 94.6 & 85.0 \\
& SRM & 0.175 & 0.361 & 0.896 & 0.948 & 0.854 & 0.837 & 0.791 & 0.479 & \textbf{96.8} & \textbf{98.9} \\
& Proc & \textbf{0.187} & \textbf{0.366} & 0.896 & 0.947 & 0.853 & 0.808 & 0.794 & 0.477 & 94.6 & 95.7 \\
\midrule

\multirow{3}{*}{$5\rightarrow1$}
& MED-VAE & 0.156 & 0.345 & \textbf{0.851} & \textbf{0.939} & \textbf{0.911} & \textbf{0.837} & \textbf{0.736} & \textbf{0.439} & \textbf{85.0} & 68.8 \\
& SRM & \textbf{0.174} & \textbf{0.355} & 0.829 & 0.925 & 0.877 & 0.808 & 0.793 & 0.476 & 77.4 & \textbf{79.6} \\
& Proc & 0.156 & 0.349 & 0.822 & 0.920 & 0.877 & 0.819 & 0.795 & 0.474 & 80.7 & 77.9 \\
\midrule

\multirow{3}{*}{$7\rightarrow1$}
& MED-VAE & \textbf{0.138} & 0.323 & \textbf{0.838} & \textbf{0.932} & \textbf{0.868} & \textbf{0.849} & \textbf{0.776} & \textbf{0.460} & \textbf{73.1} & \textbf{67.4} \\
& SRM & 0.126 & 0.334 & 0.795 & 0.866 & 0.755 & 0.763 & 0.858 & 0.518 & 61.8 & 59.1 \\
& Proc & 0.118 & \textbf{0.336} & 0.798 & 0.886 & 0.757 & 0.735 & 0.867 & 0.528 & 61.3 & 62.4 \\
\midrule

\multirow{3}{*}{$1\rightarrow2$}
& MED-VAE & 0.206 & 0.357 & \textbf{0.899} & \textbf{0.960} & \textbf{0.928} & \textbf{0.882} & \textbf{0.714} & \textbf{0.422} & 92.8 & 87.1 \\
& SRM & \textbf{0.221} & 0.371 & 0.886 & 0.959 & 0.873 & 0.833 & 0.760 & 0.448 & \textbf{96.8} & \textbf{94.6} \\
& Proc & \textbf{0.224} & \textbf{0.373} & 0.911 & 0.959 & 0.848 & 0.805 & 0.789 & 0.455 & 97.8 & 96.8 \\
\midrule

\multirow{3}{*}{$5\rightarrow2$}
& MED-VAE & \textbf{0.146} & 0.345 & 0.839 & \textbf{0.950} & \textbf{0.915} & \textbf{0.863} & \textbf{0.734} & \textbf{0.445} & \textbf{88.2} & 73.1 \\
& SRM & 0.143 & 0.350 & \textbf{0.842} & 0.927 & 0.873 & 0.831 & 0.796 & 0.478 & 80.6 & \textbf{78.5} \\
& Proc & 0.156 & \textbf{0.354} & 0.850 & 0.952 & 0.872 & 0.852 & 0.789 & 0.469 & 79.6 & 76.3 \\
\midrule

\multirow{3}{*}{$7\rightarrow2$}
& MED-VAE & \textbf{0.130} & 0.328 & \textbf{0.814} & \textbf{0.919} & \textbf{0.871} & \textbf{0.822} & \textbf{0.752} & \textbf{0.454} & \textbf{76.3} & \textbf{65.6} \\
& SRM & 0.113 & \textbf{0.350} & 0.760 & 0.873 & 0.755 & 0.739 & 0.849 & 0.521 & 67.7 & 63.4 \\
& Proc & 0.103 & 0.347 & 0.753 & 0.850 & 0.754 & 0.711 & 0.858 & 0.532 & 57.0 & 59.1 \\
\midrule

\multirow{3}{*}{$1\rightarrow5$}
& MED-VAE & \textbf{0.170} & 0.344 & 0.883 & \textbf{0.961} & \textbf{0.929} & \textbf{0.877} & \textbf{0.711} & \textbf{0.423} & 87.1 & 80.6 \\
& SRM & 0.145 & 0.357 & \textbf{0.890} & 0.929 & 0.848 & 0.809 & 0.808 & 0.481 & \textbf{93.5} & \textbf{88.2} \\
& Proc & 0.137 & \textbf{0.357} & 0.864 & 0.929 & 0.840 & 0.771 & 0.820 & 0.486 & 91.4 & 87.1 \\
\midrule

\multirow{3}{*}{$2\rightarrow5$}
& MED-VAE & \textbf{0.194} & 0.358 & \textbf{0.895} & \textbf{0.959} & \textbf{0.922} & \textbf{0.896} & \textbf{0.728} & \textbf{0.430} & \textbf{94.6} & 83.9 \\
& SRM & 0.169 & \textbf{0.373} & 0.881 & 0.941 & 0.850 & 0.810 & 0.795 & 0.481 & 90.3 & \textbf{87.1} \\
& Proc & 0.170 & 0.370 & 0.877 & 0.941 & 0.794 & 0.787 & 0.816 & 0.496 & 87.1 & \textbf{89.2} \\
\midrule

\multirow{3}{*}{$7\rightarrow5$}
& MED-VAE & \textbf{0.134} & 0.320 & \textbf{0.825} & \textbf{0.921} & \textbf{0.906} & \textbf{0.828} & \textbf{0.759} & \textbf{0.452} & \textbf{71.0} & \textbf{65.0} \\
& SRM & 0.121 & \textbf{0.345} & 0.794 & 0.848 & 0.746 & 0.723 & 0.848 & 0.517 & 57.0 & 63.4 \\
& Proc & 0.101 & 0.339 & 0.749 & 0.809 & 0.703 & 0.709 & 0.874 & 0.531 & 59.1 & 57.0 \\
\midrule

\multirow{3}{*}{$1\rightarrow7$}
& MED-VAE & \textbf{0.197} & 0.350 & \textbf{0.911} & \textbf{0.952} & \textbf{0.888} & \textbf{0.856} & \textbf{0.752} & \textbf{0.436} & 95.7 & 82.8 \\
& SRM & 0.159 & \textbf{0.365} & 0.872 & 0.945 & 0.856 & 0.819 & 0.814 & 0.487 & \textbf{96.8} & \textbf{93.5} \\
& Proc & 0.174 & 0.366 & 0.886 & 0.943 & 0.866 & 0.811 & 0.798 & 0.473 & 94.6 & \textbf{95.7} \\
\midrule

\multirow{3}{*}{$2\rightarrow7$}
& MED-VAE & \textbf{0.188} & 0.352 & 0.892 & \textbf{0.949} & \textbf{0.905} & \textbf{0.846} & \textbf{0.748} & \textbf{0.432} & 92.5 & 86.0 \\
& SRM & 0.183 & \textbf{0.361} & \textbf{0.896} & 0.945 & 0.847 & 0.814 & 0.795 & 0.460 & \textbf{93.5} & \textbf{90.3} \\
& Proc & 0.173 & 0.358 & 0.897 & 0.934 & 0.858 & 0.800 & 0.801 & 0.467 & 92.5 & 93.5 \\
\midrule

\multirow{3}{*}{$5\rightarrow7$}
& MED-VAE & 0.142 & 0.342 & 0.849 & \textbf{0.942} & \textbf{0.883} & \textbf{0.862} & \textbf{0.747} & \textbf{0.453} & \textbf{87.1} & 69.9 \\
& SRM & \textbf{0.143} & \textbf{0.354} & \textbf{0.862} & 0.932 & 0.842 & 0.812 & 0.781 & 0.475 & 83.9 & \textbf{76.3} \\
& Proc & 0.127 & 0.344 & 0.837 & 0.924 & 0.875 & 0.820 & 0.795 & 0.482 & 79.6 & 73.1 \\
\midrule\midrule

\multirow{3}{*}{\textbf{Mean}}
& MED-VAE & \textbf{0.166} & 0.344 & \textbf{0.867} & \textbf{0.946} & \textbf{0.904} & \textbf{0.856} & \textbf{0.742} & \textbf{0.440} & \textbf{86.5} & 76.3 \\
& SRM & 0.156 & \textbf{0.355} & 0.859 & 0.928 & 0.835 & 0.800 & 0.806 & 0.485 & 83.1 & \textbf{81.1} \\
& Proc & 0.152 & \textbf{0.355} & 0.845 & 0.916 & 0.825 & 0.786 & 0.816 & 0.489 & 81.3 & 80.3 \\
\bottomrule
\end{tabular}
\end{table*}

\begin{table*}[h]
\caption{MED-VAE 32D vs SRM 32D decoding performance (with default Refinement; tp=13, CFG=5). Bold indicates the better-performing method per pair. Higher is better ($\uparrow$) for all metrics except Eff and SwAV ($\downarrow$).}
\label{tab:vae_srm_32d_refine}
\centering
\scriptsize
\begin{tabular}{ll cccccccc cc}
\toprule
Pair & Method
& PixCorr$\uparrow$
& SSIM$\uparrow$
& Alex2$\uparrow$
& Alex5$\uparrow$
& Incep$\uparrow$
& CLIP$\uparrow$
& Eff$\downarrow$
& SwAV$\downarrow$
& Fwd\%$\uparrow$
& Bwd\%$\uparrow$ \\
\midrule

\multirow{2}{*}{$2\rightarrow1$}
& MED-VAE & \textbf{0.193} & 0.370 & \textbf{0.860} & \textbf{0.929} & \textbf{0.856} & \textbf{0.851} & \textbf{0.787} & \textbf{0.456} & \textbf{85.0} & 65.6 \\
& SRM & 0.144 & \textbf{0.371} & 0.805 & 0.906 & 0.764 & 0.789 & 0.873 & 0.520 & 78.5 & \textbf{71.0} \\
\midrule

\multirow{2}{*}{$5\rightarrow1$}
& MED-VAE & \textbf{0.168} & \textbf{0.359} & \textbf{0.840} & \textbf{0.935} & \textbf{0.847} & \textbf{0.826} & \textbf{0.777} & \textbf{0.452} & \textbf{79.2} & \textbf{53.8} \\
& SRM & 0.124 & 0.359 & 0.776 & 0.873 & 0.792 & 0.810 & 0.841 & 0.511 & 64.5 & 51.6 \\
\midrule

\multirow{2}{*}{$7\rightarrow1$}
& MED-VAE & \textbf{0.122} & 0.340 & \textbf{0.809} & \textbf{0.885} & \textbf{0.808} & \textbf{0.811} & \textbf{0.816} & \textbf{0.482} & \textbf{59.1} & \textbf{53.8} \\
& SRM & 0.059 & \textbf{0.358} & 0.752 & 0.817 & 0.706 & 0.701 & 0.894 & 0.549 & 45.2 & 35.5 \\
\midrule

\multirow{2}{*}{$1\rightarrow2$}
& MED-VAE & \textbf{0.210} & \textbf{0.379} & \textbf{0.880} & \textbf{0.958} & \textbf{0.902} & \textbf{0.904} & \textbf{0.727} & \textbf{0.412} & \textbf{90.3} & \textbf{77.8} \\
& SRM & 0.159 & 0.370 & 0.823 & 0.889 & 0.755 & 0.762 & 0.871 & 0.511 & 78.5 & 69.9 \\
\midrule

\multirow{2}{*}{$5\rightarrow2$}
& MED-VAE & \textbf{0.120} & 0.364 & \textbf{0.834} & \textbf{0.928} & \textbf{0.840} & \textbf{0.840} & \textbf{0.803} & \textbf{0.468} & \textbf{71.0} & \textbf{62.4} \\
& SRM & 0.087 & \textbf{0.370} & 0.746 & 0.886 & 0.811 & 0.785 & 0.841 & 0.497 & 61.3 & 56.3 \\
\midrule

\multirow{2}{*}{$7\rightarrow2$}
& MED-VAE & \textbf{0.130} & 0.348 & \textbf{0.779} & \textbf{0.872} & \textbf{0.830} & \textbf{0.810} & \textbf{0.812} & \textbf{0.477} & \textbf{61.3} & \textbf{58.1} \\
& SRM & 0.071 & \textbf{0.351} & 0.719 & 0.828 & 0.685 & 0.718 & 0.902 & 0.547 & 41.2 & 33.3 \\
\midrule

\multirow{2}{*}{$1\rightarrow5$}
& MED-VAE & \textbf{0.195} & 0.373 & \textbf{0.903} & \textbf{0.960} & \textbf{0.924} & \textbf{0.907} & \textbf{0.680} & \textbf{0.399} & \textbf{90.3} & \textbf{74.2} \\
& SRM & 0.154 & \textbf{0.375} & 0.796 & 0.886 & 0.792 & 0.782 & 0.861 & 0.515 & 76.3 & 71.0 \\
\midrule

\multirow{2}{*}{$2\rightarrow5$}
& MED-VAE & \textbf{0.163} & 0.371 & \textbf{0.836} & \textbf{0.926} & \textbf{0.865} & \textbf{0.844} & \textbf{0.769} & \textbf{0.442} & \textbf{75.3} & \textbf{68.8} \\
& SRM & 0.118 & \textbf{0.381} & 0.832 & 0.902 & 0.748 & 0.741 & 0.865 & 0.503 & 69.9 & 55.9 \\
\midrule

\multirow{2}{*}{$7\rightarrow5$}
& MED-VAE & \textbf{0.122} & 0.344 & \textbf{0.838} & \textbf{0.891} & \textbf{0.835} & \textbf{0.826} & \textbf{0.792} & \textbf{0.477} & \textbf{64.5} & \textbf{55.9} \\
& SRM & 0.055 & \textbf{0.350} & 0.753 & 0.830 & 0.690 & 0.696 & 0.894 & 0.552 & 36.6 & 34.8 \\
\midrule

\multirow{2}{*}{$1\rightarrow7$}
& MED-VAE & \textbf{0.211} & \textbf{0.382} & \textbf{0.894} & \textbf{0.948} & \textbf{0.882} & \textbf{0.896} & \textbf{0.712} & \textbf{0.406} & \textbf{89.3} & \textbf{74.2} \\
& SRM & 0.154 & 0.378 & 0.785 & 0.872 & 0.774 & 0.770 & 0.866 & 0.523 & 79.0 & 69.9 \\
\midrule

\multirow{2}{*}{$2\rightarrow7$}
& MED-VAE & \textbf{0.186} & 0.369 & \textbf{0.856} & \textbf{0.914} & \textbf{0.865} & \textbf{0.893} & \textbf{0.754} & \textbf{0.440} & \textbf{81.7} & \textbf{65.6} \\
& SRM & 0.144 & \textbf{0.387} & 0.807 & 0.889 & 0.778 & 0.786 & 0.842 & 0.504 & 68.8 & 62.4 \\
\midrule

\multirow{2}{*}{$5\rightarrow7$}
& MED-VAE & \textbf{0.134} & 0.353 & \textbf{0.822} & \textbf{0.923} & \textbf{0.862} & \textbf{0.820} & \textbf{0.787} & \textbf{0.454} & \textbf{76.3} & \textbf{62.4} \\
& SRM & 0.120 & \textbf{0.372} & 0.760 & 0.864 & 0.798 & 0.810 & 0.839 & 0.502 & 69.9 & 55.9 \\
\midrule\midrule

\multirow{2}{*}{\textbf{Mean}}
& MED-VAE & \textbf{0.163} & 0.363 & \textbf{0.846} & \textbf{0.922} & \textbf{0.860} & \textbf{0.852} & \textbf{0.768} & \textbf{0.447} & \textbf{77.0} & \textbf{64.4} \\
& SRM & 0.116 & \textbf{0.369} & 0.779 & 0.870 & 0.754 & 0.764 & 0.867 & 0.517 & 64.1 & 55.6 \\
\bottomrule
\end{tabular}
\end{table*}

\begin{table*}[h]
\caption{MED-VAE 512D vs SRM 512D decoding performance (with default Refinement; tp=13, CFG=5). Bold indicates the better-performing method per pair. Higher is better for all metrics except Eff and SwAV (lower is better).}
\label{tab:vae_srm_512d_refine}
\centering
\scriptsize
\begin{tabular}{ll cccccccc cc}
\toprule
Pair & Method 
& PixCorr$\uparrow$ 
& SSIM$\uparrow$ 
& Alex2$\uparrow$ 
& Alex5$\uparrow$ 
& Incep$\uparrow$ 
& CLIP$\uparrow$ 
& Eff$\downarrow$ 
& SwAV$\downarrow$ 
& Fwd\%$\uparrow$ 
& Bwd\%$\uparrow$ \\
\midrule

\multirow{2}{*}{$2\rightarrow1$}
& MED-VAE & \textbf{0.215} & \textbf{0.382} & \textbf{0.888} & 0.948 & \textbf{0.915} & \textbf{0.896} & \textbf{0.752} & \textbf{0.435} & 93.9 & 85.0 \\
& SRM & 0.214 & 0.377 & 0.882 & 0.948 & 0.827 & 0.847 & 0.794 & 0.463 & \textbf{96.8} & \textbf{98.9} \\
\midrule

\multirow{2}{*}{$5\rightarrow1$}
& MED-VAE & 0.163 & 0.357 & \textbf{0.862} & \textbf{0.940} & \textbf{0.881} & \textbf{0.887} & \textbf{0.742} & \textbf{0.433} & \textbf{85.0} & 68.8 \\
& SRM & \textbf{0.186} & \textbf{0.371} & 0.828 & 0.915 & 0.840 & 0.863 & 0.782 & 0.463 & 77.4 & \textbf{79.6} \\
\midrule

\multirow{2}{*}{$7\rightarrow1$}
& MED-VAE & \textbf{0.128} & 0.349 & \textbf{0.842} & \textbf{0.909} & \textbf{0.848} & \textbf{0.864} & \textbf{0.770} & \textbf{0.452} & \textbf{73.1} & \textbf{68.8} \\
& SRM & 0.104 & \textbf{0.364} & 0.787 & 0.858 & 0.773 & 0.735 & 0.848 & 0.516 & 61.8 & 59.1 \\
\midrule

\multirow{2}{*}{$1\rightarrow2$}
& MED-VAE & 0.204 & 0.383 & 0.903 & \textbf{0.967} & \textbf{0.901} & \textbf{0.898} & \textbf{0.709} & \textbf{0.404} & 92.5 & 87.1 \\
& SRM & \textbf{0.212} & \textbf{0.392} & \textbf{0.905} & 0.951 & 0.835 & 0.805 & 0.802 & 0.460 & \textbf{96.8} & \textbf{94.6} \\
\midrule

\multirow{2}{*}{$5\rightarrow2$}
& MED-VAE & \textbf{0.130} & 0.367 & \textbf{0.848} & \textbf{0.937} & \textbf{0.887} & \textbf{0.862} & \textbf{0.760} & \textbf{0.443} & \textbf{88.2} & 73.1 \\
& SRM & 0.120 & \textbf{0.382} & 0.820 & 0.916 & 0.840 & 0.843 & 0.804 & 0.478 & 80.6 & \textbf{78.5} \\
\midrule

\multirow{2}{*}{$7\rightarrow2$}
& MED-VAE & \textbf{0.140} & 0.361 & \textbf{0.825} & \textbf{0.904} & \textbf{0.873} & \textbf{0.848} & \textbf{0.766} & \textbf{0.447} & \textbf{76.3} & \textbf{65.6} \\
& SRM & 0.116 & \textbf{0.377} & 0.775 & 0.875 & 0.754 & 0.765 & 0.849 & 0.518 & 67.7 & 63.4 \\
\midrule

\multirow{2}{*}{$1\rightarrow5$}
& MED-VAE & \textbf{0.202} & 0.370 & \textbf{0.908} & \textbf{0.963} & \textbf{0.895} & \textbf{0.903} & \textbf{0.713} & \textbf{0.417} & 87.1 & 80.6 \\
& SRM & 0.174 & \textbf{0.376} & 0.887 & 0.946 & 0.838 & 0.795 & 0.783 & 0.464 & \textbf{93.5} & \textbf{88.2} \\
\midrule

\multirow{2}{*}{$2\rightarrow5$}
& MED-VAE & \textbf{0.191} & 0.383 & \textbf{0.885} & \textbf{0.959} & \textbf{0.900} & \textbf{0.897} & \textbf{0.729} & \textbf{0.421} & \textbf{94.6} & 83.9 \\
& SRM & 0.180 & \textbf{0.392} & 0.876 & 0.934 & 0.849 & 0.802 & 0.788 & 0.465 & 90.3 & \textbf{87.1} \\
\midrule

\multirow{2}{*}{$7\rightarrow5$}
& MED-VAE & \textbf{0.133} & 0.358 & \textbf{0.858} & \textbf{0.926} & \textbf{0.872} & \textbf{0.867} & \textbf{0.731} & \textbf{0.432} & \textbf{71.0} & \textbf{64.5} \\
& SRM & 0.088 & \textbf{0.368} & 0.785 & 0.841 & 0.692 & 0.762 & 0.827 & 0.507 & 57.0 & 63.4 \\
\midrule

\multirow{2}{*}{$1\rightarrow7$}
& MED-VAE & \textbf{0.217} & 0.374 & \textbf{0.896} & \textbf{0.959} & \textbf{0.883} & \textbf{0.874} & \textbf{0.730} & \textbf{0.421} & 95.7 & 82.8 \\
& SRM & 0.200 & \textbf{0.388} & 0.890 & 0.941 & 0.841 & 0.814 & 0.795 & 0.470 & \textbf{96.8} & \textbf{93.5} \\
\midrule

\multirow{2}{*}{$2\rightarrow7$}
& MED-VAE & \textbf{0.205} & 0.377 & \textbf{0.891} & \textbf{0.944} & \textbf{0.883} & \textbf{0.869} & \textbf{0.731} & \textbf{0.425} & 92.5 & 86.0 \\
& SRM & 0.174 & \textbf{0.382} & 0.870 & 0.922 & 0.835 & 0.829 & 0.794 & 0.462 & \textbf{93.5} & \textbf{90.3} \\
\midrule

\multirow{2}{*}{$5\rightarrow7$}
& MED-VAE & \textbf{0.149} & 0.356 & \textbf{0.833} & \textbf{0.941} & \textbf{0.903} & \textbf{0.860} & \textbf{0.739} & \textbf{0.430} & \textbf{87.1} & 69.9 \\
& SRM & 0.128 & \textbf{0.372} & 0.816 & 0.914 & 0.847 & 0.846 & 0.783 & 0.462 & 83.9 & \textbf{76.3} \\
\midrule\midrule

\multirow{2}{*}{\textbf{Mean}}
& MED-VAE & \textbf{0.173} & 0.368 & \textbf{0.870} & \textbf{0.941} & \textbf{0.887} & \textbf{0.877} & \textbf{0.739} & \textbf{0.430} & \textbf{86.4} & 76.3 \\
& SRM & 0.158 & \textbf{0.379} & 0.844 & 0.919 & 0.813 & 0.809 & 0.805 & 0.477 & 83.1 & \textbf{81.1} \\
\bottomrule
\end{tabular}
\end{table*}

\begin{figure}[!t]
    \centering    \includegraphics[width=0.99\textwidth]{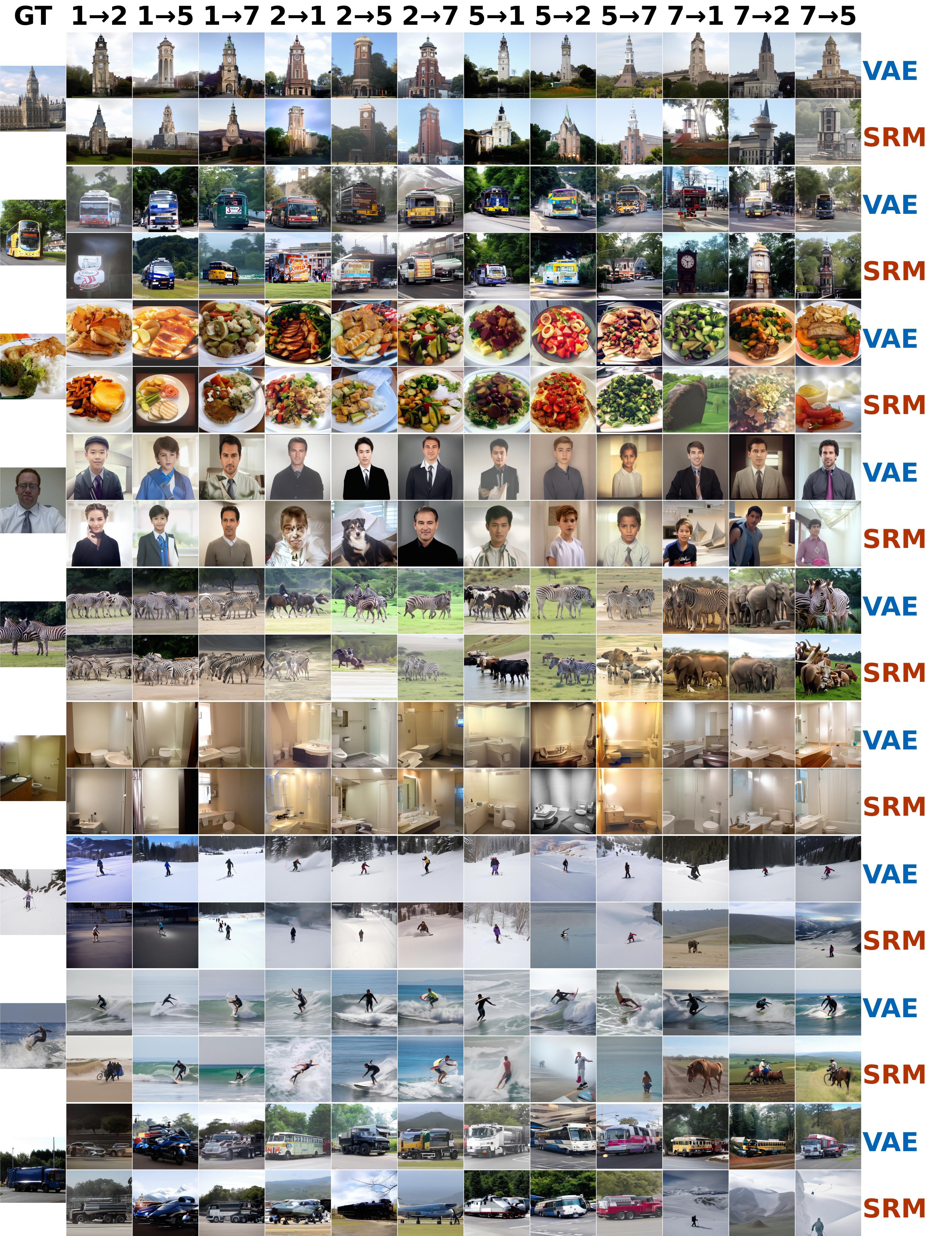}
    \label{fig:reconstr_appendix}
    \caption{\small\textbf{Cross-subject image reconstruction comparison between VAE and SRM alignment methods.} Each block shows reconstructions for a single stimulus GT (ground truth shown in the leftmost column) across all 12 subject transfer pairs (i→j). For each stimulus, the top row shows VAE reconstructions and the bottom row shows SRM reconstructions. Both methods use 512-dimensional latent representations with nearest-neighbor refinement (k=5)}
    \label{fig:reconstructed_images}
\end{figure}

\end{document}